\documentclass[11pt,a4paper]{article}
\usepackage{float}
\usepackage{multirow}
\usepackage{graphics}
\usepackage[cp1252,utf8]{inputenc}
\usepackage{hyperref}
\usepackage{graphicx}
\usepackage{wrapfig}
\usepackage{amsmath}
\usepackage{amsfonts}
\usepackage{amssymb}
\usepackage{relsize} 
\usepackage[top=1in, bottom=2cm, left=1.5cm, right=1.5cm]{geometry}
\usepackage{authblk}
\usepackage{physics}
\NewDocumentCommand{\tens}{t_}
{%
	\IfBooleanTF{#1}
	{\tensop}
	{\otimes}%
}
\NewDocumentCommand{\tensop}{m}
{%
	\mathbin{\mathop{\otimes}\displaylimits_{#1}}%
}
\usepackage{changepage} 

\usepackage{afterpage}

\usepackage{placeins}

\usepackage{wrapfig}
\usepackage{cite} 
\usepackage[nottoc,notlof,notlot]{tocbibind} 

\title{\bf Entanglement wedge cross section for noncommutative Yang-Mills theory}
\author[a]{\bf  Anirban Roy Chowdhury \thanks{iamanirban@bose.res.in}}
\author[b]{\bf  Ashis Saha \thanks{ashisphys18@klyuniv.ac.in}}
\author[c]{\bf Sunandan Gangopadhyay \thanks{sunandan.gangopadhyay@bose.res.in}}

\affil[a,c]{\textit{Department of Theoretical Sciences,
		S.N.~Bose National Centre for Basic Sciences,}
	\textit{JD Block, Sector-III, Salt Lake, Kolkata 700106, India}}
\affil[b]{\textit{Department of Physics, University of Kalyani, Kalyani 741235, India}}

\date{}

\begin{document}
	\maketitle
\begin{abstract}
\noindent The signature of noncommutativity on various measures of entanglement has been observed by considering the holographic dual of noncommutative super Yang-Mills theory. We have followed a systematic analytical approach in order to compute the holographic entanglement entropy corresponding to a strip like subsystem of length $l$. The relationship between the subsystem size (in dimensionless form) $\frac{l}{a}$ and the turning point (in dimensionless form) introduces a critical length scale $\frac{l_c}{a}$ which leads to three domains in the theory, namely, the deep UV domain ($l< l_c$; $au_{t}\gg 1$, $au_{t}\sim au_{b}$), deep noncommutative domain ($l> l_c,~au_{b}>au_t\gg 1$) and deep IR domain ($l> l_c,~au_t\ll 1$). This in turn means that the length scale $l_c$ distinctly points out the UV/IR mixing property of the non-local theory under consideration. We have carried out the holographic study of entanglement entropy for each of these domains by employing both analytical and numerical techniques. The broken Lorentz symmetry induced by noncommutativity has motivated us to redefine the entropic $c$-function. We have obtained the noncommutative correction to the $c$-function upto leading order in the noncommutative parameter. We have also looked at the behaviour of this quantity over all the domains of the theory. We then move on to compute the minimal cross-section area of the entanglement wedge by considering two disjoint subsystems $A$ and $B$. On the basis of $E_P = E_W$ duality, this leads to the holographic computation of the entanglement of purification. The correlation between two subsystems, namely, the holographic mutual information $I(A:B)$ has also been computed. Moreover, the computations of $E_W$ and $I(A:B)$ has been done for each of the domains in the theory. We have then briefly discussed the effect of the UV cut-off on the IR behaviours of these quantities.
Finally, we consider a black hole geometry with a noncommutative parameter and study the influence of both noncommutativity and finite temperature on the various measures of quantum entanglement.
\end{abstract}
\section{Introduction}
The study of information theoretic quantities has proven to be a crucial tool to understand the fundamental aspects of quantum mechanics and quantum field theory. Among various entanglement measures, the entanglement entropy (EE) is one of the most useful physical quantity with a very simple definition. The EE is a good measure of entanglement for pure states and it is defined as the von-Neumann entropy of the reduced density matrix. However, it is not a suitable measure of entanglement for mixed states as it measures both quantum and classical correlations. In order to resolve this, the concepts of entanglement of purification, reflected entropy and entanglement negativity were introduced. In this paper we will focus on the idea of entanglement of purification (EoP) \cite{Terhal_2002} in the holographic set up. The EoP can be defined as follows. If $\rho_{AB}$ represents a density matrix corresponding to a mixed state in Hilbert spcae $\mathcal{H} =\mathcal{H}_A \tens \mathcal{H}_B$, then the process of purification suggests that a pure state $\ket{\psi}$ can be computed from $\rho_{AB}$ by adding auxiliary degrees of freedom to the total Hilbert space $\mathcal{H}$. This obtained pure state $\ket{\psi}$ is often denoted as one of the possible purifications of $\rho_{AB}$. It is worth mentioning that the process of purification is not unique. Further, the measure for entanglement in this set up which is denoted as the EoP has the following definition
\cite{Terhal_2002}
\begin{eqnarray}\label{EoP}
E_P(\rho_{AB})\equiv E_P(A,B) = \mathop{min}_{\ket{\psi}}S(\rho_{AA^{\prime}});~\rho_{AA^{\prime}} = tr_{BB^{\prime}}\ket{\psi}\bra{\psi}
\end{eqnarray}
where the minimization is taken over any state $\ket{\psi}$ with $\rho_{AB} = tr_{A^{\prime}B^{\prime}}\ket{\psi}\bra{\psi}$ being held constant. However, in the field theoretic scenario, the computations of this measure become quite challenging. This was resolved by the gauge/gravity duality upto a satisfactory level. The gauge/gravity duality \cite{Maldacena:1997re,Witten:1998qj,Aharony:1999ti} has related these mentioned entanglement measures of the boundary field theory to certain geometric regions in the bulk gravity theory \cite{Ryu:2006bv,Ryu:2006ef,Nishioka:2009un}. The gauge/gravity duality led holographic computation of EE is known as the Ryu-Takayanagi (RT) prescription. The RT prescription relates the area of a codimension-2 static minimal surface in the bulk theory to the von-Neumann entropy (EE) of the reduced density matrix at the boundary QFT \cite{Ryu:2006bv,Ryu:2006ef,Nishioka:2009un}. This can be expressed as
\begin{eqnarray}
S_{EE}(A) = \frac{\mathrm{Area}(\Gamma_A^{min})}{4G_N}
\end{eqnarray}
where $S_{EE}(A)$ denotes the EE of the subsystem $A$ at the boundary QFT. On the other hand, the holographic analogy of EoP has been suggested to be the minimal cross section of the entanglement wedge (EWCS) $E_W(A,B)$ \cite{Takayanagi:2017knl,Nguyen:2017yqw}. It has been observed that both $E_P(A,B)$ and $E_W(A,B)$ satisfy the following properties \cite{Takayanagi:2017knl}
\begin{eqnarray}\label{prop}
&&~E_P(A,B) = S_{EE}(A)=S_{EE}(B);~\rho_{AB}^2=\rho_{AB},\nonumber\\
&&~\frac{1}{2} I(A:B) \leq E_P(A,B) \leq min\left[S_{EE}(A),S_{EE}(B)\right], \nonumber\\
&& \frac{I(A:B)+I(A:C)}{2} \leq E_P(A,B\cup C)
\end{eqnarray}
where $I(A:B)$ represents the mutual information  between two subsystems $A$ and $B$ given by
\begin{equation}\label{mi}
I(A:B)=S_{EE}(A)+S_{EE}(B)-S_{EE}(A\cup B)~.
\end{equation}
Replacing $E_{P}$ by $E_{W}$ in eq.(\ref{prop}), we obtain the inequalities for $E_{W}$. Note that the inequalities appearing in the second and third lines hold for both pure and mixed states.
It is worth mentioning that for a certain critical separation length between $A$ and $B$, the mutual information vanishes ($I(A:B)=0$) and the domain of entanglement wedge becomes disconnected. Apart from $E_W$, various different quantities have been suggested to probe mixed state correlation measures, some of them are, reflected entropy \cite{Dutta:2019gen,Chu:2019etd} and logarithmic negativity \cite{Kudler-Flam:2018qjo,Kusuki:2019zsp}. Furthermore, recently it was suggested that information associated to the EWCS can be extracted from the odd entropy $S_O(A,B)$, as $S_O(A,B)=E_W(A,B)+S_{EE}(A\cup B)$ \cite{Tamaoka:2018ned}. Due to its usefulness and interesting properties, the study of EWCS has gained appreciable amount of attention in recent times. Some of these interesting observations in this direction can be found in \cite{Espindola:2018ozt,Agon:2018lwq,Umemoto:2018jpc,Hirai:2018jwy,Bao:2019wcf,Kusuki:2019evw,Jeong:2019xdr,Umemoto:2019jlz,Harper:2019lff,Jokela:2019ebz,BabaeiVelni:2019pkw,Ghodrati:2019hnn,Kusuki:2019hcg,Boruch:2020wbe,BabaeiVelni:2020wfl,Saha:2021kwq,Camargo:2020yfv,Liu:2021rks,Sahraei:2021wqn}.\\ 
\noindent On the other hand, noncommutativity of spacetime is a very unique concept which has appeared in various areas of physics and mathematics. The fundamental philosophy of noncommutativity states that spacetime coordinates do not commute and satisfies the following relation \cite{Snyder}
\begin{equation}\label{nc}
[x_{i},x_{j}]=\mathnormal{i}\vartheta_{ij}
\end{equation}
where $\vartheta_{ij}$ is anti-symmetric in $i,j$; where $i,j$ can take all possible values. Since in this work we shall consider noncommutativity only in a plane, hence only one coordinate pair of $\vartheta_{ij}$ will be non-zero.
In this paper we consider the noncommutative generalization of the Yang-Mills theory. In order to make a gauge theory noncommutative one needs to deform the products of ordinary functions in the following form 
\begin{equation}
(f\star g)(x)= e^{(i/2)\vartheta^{ij}\frac{\partial}{\partial x^{i}}\frac{\partial}{\partial x^{j}}}f(x)g(y)\arrowvert_{y=x}~;~ i,j=2,3~;~x_{2}=x,~x_{3}=y
\end{equation}
where $f(x)$ and $g(y)$ are ordinary functions. The above product is denoted as the Moyal-Weyl product (star product) \cite{Minwalla:1999px,Armoni:2000xr,Szabo:2001kg}. The motativations to study this type of star product deformed gauge theory are their non-local nature and UV/IR mixing property \cite{Minwalla:1999px,Armoni:2000xr}. In the context of string theory, noncommutative gauge theory arises as the low-energy limiting theories of D-branes with non-vanishing NS-NS $B$-field background \cite{Connes:1997cr,Seiberg:1999vs,Ardalan:1998ce}. Some studies related to the emergence of noncommutativity due to the presence of a background
NS two-form field $B_{\mu\nu}$ can be found in \cite{Chakraborty:2006yj,Gangopadhyay:2007ne,Chatterjee:2008su}. In this paper we probe the effect of this noncommutative deformation of a gauge theory on various entanglement measures. In particular we start from a gravity dual spacetime geometry dual to a noncommutative (NC) Yang-Mills theory. With this geometry, we compute the holographic entanglement entropy (HEE), entanglement wedge-cross section (EWCS) and mutual information. We compute all of these quantities in different domains of the theory, and compare the numerical and the analytical results. Our motivation is to investigate the effect of the NC parameter on the above information theoretic quantities. Furthermore, we also study the effect of noncommutativity on the entropic $c$-function. Some previous observation of various entanglement measures for NC Yang-Mills can be found in \cite{Fischler:2013gsa,Karczmarek:2013xxa,Fischler:2018kwt,Nakajima:2020rrj}.\\
\noindent The paper is organized as follows. In section \eqref{Sec1}, we give a short description of the holographic dual of noncommutative Yang-Mills theory. The holographic computation of EE has been carried out in section \eqref{Sec2} and the computation of entropic $c$-function has been shown in section \eqref{newsection}. In section \eqref{Sec3} we compute the minimal cross-section of the entanglement wedge and study the effect of noncommutativity on it. The finite temperature computations has been done in section \eqref{Sec4} and section \eqref{Sec5}. We summarize our findings and conclude in section \eqref{Sec6}.
\section{Dual description of noncommutative Yang-Mills theory}\label{Sec1}
In \cite{Seiberg:1999vs}, it was
shown that the non-zero NS-NS $B$-field leads to noncommutative space on the $D$-brane which decouples from the closed string excitations. The $B$-field is introduced by performing a $T$-duality in a particular direction while the other directions are compactified on a torus. In \cite{Hashimoto:1999ut,Maldacena:1999mh}, a stack of $D3$-branes with non-zero $B$-field (in a certain plane) was considered and it was shown that at a particular decoupling limit, a holographic dual of $SU(N)$ noncommutative super Yang-Mills theory exists. This type IIB gravity dual is described by the following metric in the string frame \cite{Hashimoto:1999ut,Maldacena:1999mh}
\begin{eqnarray}\label{NCYM}
ds^{2}&=&R^{2}\bigg[-u^{2}dt^{2}+u^{2}dx_{1}^{2}+u^{2}h(u)(dx_{2}^{2}+dx_{3}^{2})+\frac{du^{2}}{u^{2}}\bigg]+R^{2}d\Omega_{5}^{2}
\end{eqnarray}
where $h(u)=\frac{1}{1+a^{4}u^{4}}$ and $a=\lambda^{1/4}\sqrt{\vartheta}$ is the renormalized noncommutative scale or the NC parameter. The NC parameter is non-zero only in the $x_2-x_3$ plane with the commutator $\left[x_2,x_3\right]=i\vartheta$. The non-vanishing dilaton profile is specified as $e^{2\Phi}=g_s^2h(u)$ where $g_s$ is the string coupling. The t'Hooft coupling constant is related with the $AdS$ radius as $\sqrt{\lambda}=\frac{R^2}{\alpha^{\prime}}$ where $\alpha^{\prime}$ is the string tension. Further the only non-vanishing component of the NS-NS $B$-field reads $B_{23} = R^2a^2u^4h(u)$.\\
We would like to make a comment now. The $(x_{2},x_{3})$-plane collapses in the UV limit, that is, as $u\rightarrow \infty$. Hence it is necessary to introduce a UV cutoff. We shall see subsequently that the introduction of this cutoff leads to the presence of a critical length, and the study of entanglement of regions smaller than this critical length needs to be done carefully. For cylindrical entangling regions, the necessity of the cutoff is the following.
Without the cutoff, all bulk surfaces would correspond to the same boundary region in the collapsing $(x_{2},x_{3})$-plane \cite{Fischler:2013gsa}.
\section{Holographic computation of entanglement entropy and UV/IR mixing}\label{Sec2}
We start our analysis by considering a strip like subsystem, namely, subsystem $A$. The subsystem is specified by the volume $V_{sub} = L^2l$, where $-\frac{l}{2}\leq x_2\leq \frac{l}{2}$ and $x_1,x_3 \in \left[-L,L\right]$ with $L\rightarrow \infty$. Further, we assume that the widths along $x_1$ and $x_3$ are fixed and only the width along $x_2$ is allowed to vary. This particular choice has been made in order to probe the effect of noncommutativity on the EE. One can also make the choice $-\frac{l}{2}\leq x_1\leq \frac{l}{2}$ and $x_2,x_3 \in \left[-L,L\right]$, for which the effect of noncommutativity does not influence the computed result of EE. We choose the parametrization $u=u(x_2)$ in order to compute the surface area of the co-dimension one RT surface $\Gamma_A^{min}$. On the other hand it is to be noted that the metric (\ref{NCYM}) is given in the 10-dimensional string frame with a non-vanishing dilaton however the calculation is to be done in the Einstein frame. In order to resolve this, we use the following transformation
\begin{eqnarray}
g_{\mu\nu}^E \rightarrow e^{-\frac{\phi}{2}} g_{\mu\nu}^S~.
\end{eqnarray}
By using the above transformation, we obtain $\sqrt{g_8^E} = e^{-2\phi}\sqrt{g_8^S}$. We now use this fact to write down the generalized RT formula for 10-dimensional string frame
\begin{eqnarray}\label{RTF}
S_{EE}&=&\frac{Area(\Gamma_A^{min})}{4G^{(10)}}\nonumber\\
&=& \frac{1}{4G_N^{(10)}} \int d^8\xi~ e^{-2\phi}\sqrt{g_8^S}\nonumber\\
&=&\frac{2 R^8 L^2 \mathrm{Vol}(\Omega_{5})}{4g_s^2G_N^{(10)}} \int_{-l/2}^{0} u^3 \sqrt{1+\frac{{u^{\prime}}^2}{u^4 h(u)}} dx_2~~;~u^{\prime}\equiv \frac{du}{dx_{2}}
\end{eqnarray}
where $G_N^{(10)}$ is the $10d$ Newton's constant which is related with the $5d$ Newton's constant as $G_N^{(10)}=\pi^3R^5G_N^{(5)}$.
Considering the integrand in the above equation as the Lagrangian, it is easy to see that $x_{2}$ is a cyclic coordinate.
This gives rise to the conserved Hamiltonian
\begin{equation}
\mathcal{H}=-\frac{u^{3}}{\sqrt{1+\frac{{u^{\prime}}^{2}}{u^{4}h(u)}}}=constant(c)~.
\end{equation}
At the turning point $u=u_{t}$, $\frac{du}{dx_{2}}=0$. This fixes the value of the constant $c=-u_{t}^{3}$, which then results in the following differential equation 
\begin{equation}\label{e}
\frac{du}{dx_{2}}=\sqrt{u^{4}h(u)\bigg(\bigg(\frac{u}{u_{t}}\bigg)^{6}-1\bigg)}~.
\end{equation}
Now substituting eq.(\ref{e}) in eq.(\ref{RTF}) and using the boundary condition (which implements the UV cutoff)
\begin{equation}\label{bc1}
u(x_{2}=\pm \frac{l}{2})=u_{b}=\frac{1}{\epsilon}
\end{equation}
we get the dimensionless form of HEE
\begin{eqnarray}\label{EE1}
a^2S_{EE}=\frac{2 R^8 L^2 \mathrm{Vol}(\Omega_{5})}{4g_s^2G_N^{(10)}} (au_t)^2 \int_{\frac{au_t}{au_b}}^{1}\frac{\sqrt{p^{4}+(au_{t})^{4}}}{p^{5}\sqrt{1-p^{6}}} dp
\end{eqnarray}
where
$p = \frac{au_t}{au}$. On the other hand the length of the subsystem (in dimensionless form) in terms of the bulk coordinate reads
\begin{equation}\label{lgen}
\frac{l}{a}=\frac{2}{au_{t}}\int_{\frac{au_{t}}{au_{b}}}^{1} dp \frac{p\sqrt{p^{4}+(au_{t})^{4}}}{\sqrt{1-p^{6}}}~.
\end{equation}
We first compute the integral given in eq.(\ref{lgen}), in order to probe the relation between the subsystem size $l$ and turning point $u_t$. It is well-known that the UV/IR mixing property is one of the most interesting aspects of this noncommutative gauge theory. We aim to probe this complicated UV/IR mixing property by following an analytical approach. The deep IR limit is characterized by the fact $au_{t}<<1$ \cite{Barbon:2008ut}. Incorporating this condition, one can obtain the following relation
\begin{eqnarray}\label{lIR}
\bigg(\frac{l}{a}\bigg)_{deep~IR}&\approx&\frac{2}{au_{t}}\int_{0}^{1} dp \frac{p^{3}}{\sqrt{1-p^{6}}}\\ \nonumber
&=&\frac{2}{(au_{t})}\sqrt{\pi}\frac{\Gamma(2/3)}{\Gamma(1/6)}~.
\end{eqnarray}
On the other hand, the deep noncommutative (NC) limit is associated with fact $au_t\gg 1$, and $au_{t}\ll au_{b}$ \cite{Barbon:2008ut}, which leads to
\begin{eqnarray}\label{lUV}
\bigg(\frac{l}{a}\bigg)_{deep~NC}&\approx&2(au_{t})\int_{0}^{1} dp \frac{p}{\sqrt{1-p^{6}}}\\ \nonumber
&=&\frac{\sqrt{\pi}}{3}\frac{\Gamma(1/3)}{\Gamma(5/6)}(au_{t})~.	
\end{eqnarray}
From the above relations it can be observed that the deep IR limit leads to the result corresponding to the usual commutative $\mathcal{N}=4$ super Yang-Mills gauge theory ($AdS_5\times S^5$) in $3+1$-dimensions. This in turn means that one can denote the deep IR limit as the commutative limit of this theory.
The deep UV limit, on the other hand, needs to be analysed carefully as we shall now see.\\ 
Our aim is to obtain a single analytical solution which can probe the UV/IR mixing property. This can be done in the following way. For $au_{t}\le 1$, eq.(\ref{lgen}) can be written as
\begin{eqnarray}\label{lorder2}
\frac{l}{a}&=& \frac{2}{(au_t)} \left[\int_{\frac{au_{t}}{au_{b}}}^{au_t} dp \frac{p\sqrt{p^{4}+(au_{t})^{4}}}{\sqrt{1-p^{6}}}+\int_{au_t}^{1} dp \frac{p\sqrt{p^{4}+(au_{t})^{4}}}{\sqrt{1-p^{6}}}\right]~.
\end{eqnarray}
In eq.(\ref{lorder2}), we have divided the whole integral in two parts. It can be noted that for the first integral $0\leq p\leq (au_t)$ and hence $\frac{p}{(au_t)}< 1$. Similarly for the second integral $(au_t)\leq p\leq 1$ and hence $\frac{(au_t)}{p}< 1$. We can now perform a binomial expansion and keep terms upto $\mathcal{O}\left(\frac{p}{au_t}\right)^4$ in the first integral and terms upto $\mathcal{O}\left(\frac{au_t}{p}\right)^4$ in the second integral. This leads to the following expression for the subsystem size  (for $au_{t} \le 1$)
\begin{eqnarray}\label{l2}
\frac{l}{a}&\approx&\frac{\sqrt{\pi}}{2(au_t)}\frac{\Gamma(\frac{5}{3})}{\Gamma(\frac{7}{6})}-(au_{t})^{3}\ln(au_{t})+(au_t)^3\sum_{n=1}^{\infty}\frac{1}{\sqrt{\pi}}\frac{\Gamma(n+\frac{1}{2})}{\Gamma(n+1)}\frac{1}{(6n)}-\sum_{n=1}^{\infty}\frac{1}{\sqrt{\pi}}\frac{\Gamma(n+\frac{1}{2})}{\Gamma(n+1)}\frac{(au_t)^{(6n+3)}}{(6n)}\nonumber\\
&+&\left(\sum_{n=0}^{\infty}\frac{2}{\sqrt{\pi}}\frac{\Gamma(n+\frac{1}{2})}{\Gamma(n+1)}\left[\frac{1-(1/au_{b})^{6n+2}}{(6n+2)}-\frac{1}{(6n+4)}+\frac{1-(1/au_{b})^{6n+6}}{2(6n+6)}\right]\right)(au_t)^{(6n+3)}~.
\end{eqnarray}
For $au_{t}\ge 1$, the expression for $\left(\frac{l}{a}\right)$ (in eq.\eqref{lgen}) reads
\begin{eqnarray}\label{L2}
\left(\frac{l}{a}\right)&=& 2(au_t)\int_{\frac{au_t}{au_b}}^1dp\sum_{n=0}^{\infty}\sum_{m=0}^{\infty}\frac{p}{\sqrt{\pi}}\frac{\Gamma(n+\frac{1}{2})}{\Gamma(n+1)}\frac{\Gamma(\frac{3}{2})}{\Gamma(m+1)\Gamma(\frac{3}{2}-m)}p^{6n}\left(\frac{p}{au_t}\right)^{4m}\nonumber\\
&=&\sum_{n,m=0}^{\infty}\frac{\Gamma(n+\frac{1}{2})}{\Gamma(n+1)\Gamma(m+1)\Gamma(\frac{3}{2}-m)}\frac{1}{(au_{t})^{4m-1}}\frac{1}{(6n+4m+2)}\left[1-\left(\frac{au_{t}}{au_{b}}\right)^{6n+4m+2}\right]\nonumber\\
&\approx&\sum_{n=0}^{\infty}\frac{\Gamma(n+\frac{1}{2})}{\Gamma(n+1)\Gamma(\frac{3}{2})}(au_{t})\frac{1}{(6n+2)}\left[1-\left(\frac{au_{t}}{au_{b}}\right)^{6n+2}\right]\nonumber\\
&&+\sum_{n=0}^{\infty}\frac{\Gamma(n+\frac{1}{2})}{\Gamma(n+1)\Gamma(2)\Gamma(\frac{1}{2})}\frac{1}{(au_{t})^{3}}\frac{1}{(6n+6)}\left[1-\left(\frac{au_{t}}{au_{b}}\right)^{6n+6}\right]
\end{eqnarray}
where in getting the first line we have used the identities
\begin{eqnarray}
\sqrt{1+\left(\frac{p}{au_t}\right)^4}&=&\sum_{m=0}^{\infty}\frac{\Gamma(\frac{3}{2})}{\Gamma(m+1)\Gamma(\frac{3}{2}-m)}\left(\frac{p}{au_t}\right)^{4m}~;~\left(\frac{p}{au_t}<1\right)\nonumber\\
\frac{1}{\sqrt{1-p^6}}&=&\sum_{n=0}^{\infty}\frac{1}{\sqrt{\pi}}\frac{\Gamma(n+\frac{1}{2})}{\Gamma(n+1)}\left(p\right)^{6n}~;~\left(p<1\right)~.
\end{eqnarray}
It can be observed that in the limit $a\rightarrow0$, eq.(\ref{l2}) produces the result corresponding to commutative SYM theory.
\begin{equation}
\left(\frac{l}{a}\right)=\frac{2}{\sqrt{\pi}(au_{t})}\sum_{n=0}^{\infty}\frac{\Gamma(n+\frac{1}{2})}{\Gamma(n+1)}\frac{1}{(6n+4)}\left[1-\left(\frac{au_{t}}{au_{b}}\right)^{6n+4}\right]~.
\end{equation}
We now numerically compute the integral given in eq.(\ref{lgen}) and compare it with our analytically computed result given in eq.(\ref{l2}) and eq.(\ref{L2}).\\
\begin{figure}[!h]
	\begin{minipage}[t]{0.48\textwidth}
		\centering\includegraphics[width=\textwidth]{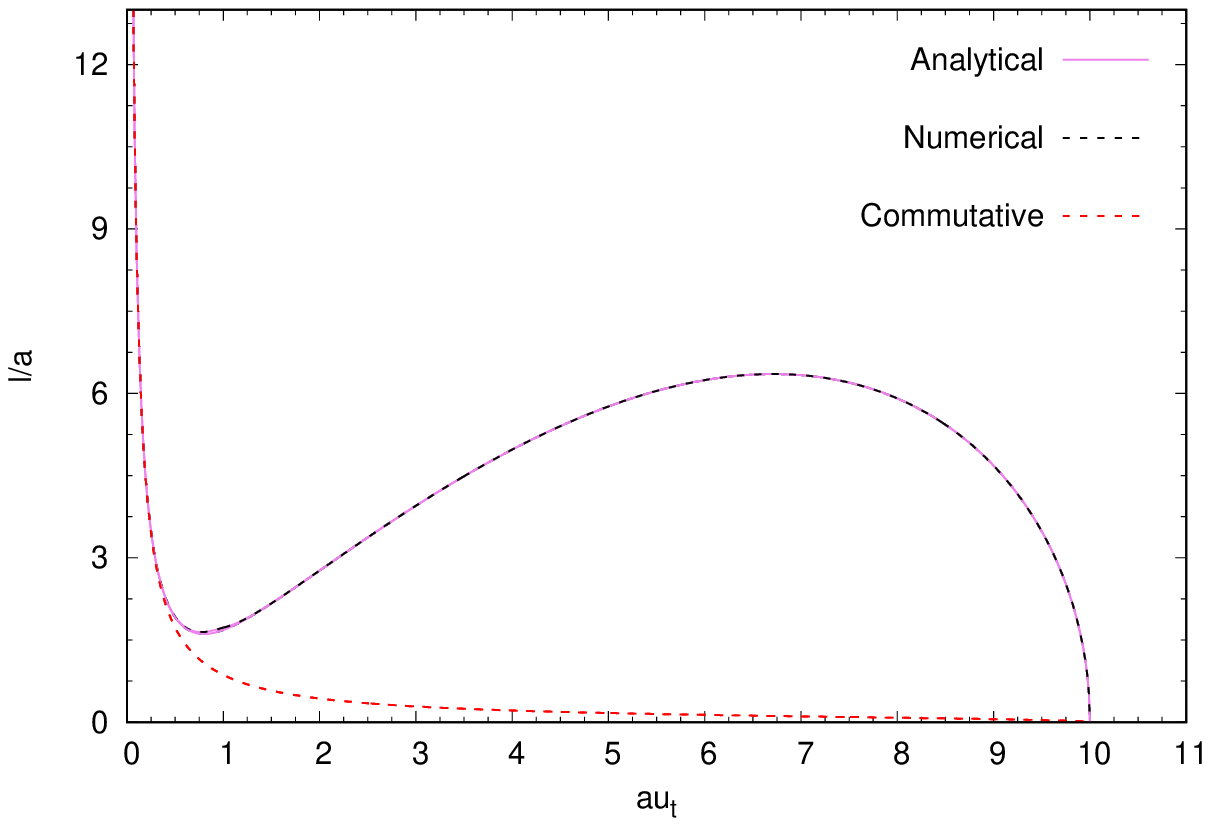}\\
		{\footnotesize  $au_b=10$}
	\end{minipage}\hfill
	\begin{minipage}[t]{0.48\textwidth}
		\centering\includegraphics[width=\textwidth]{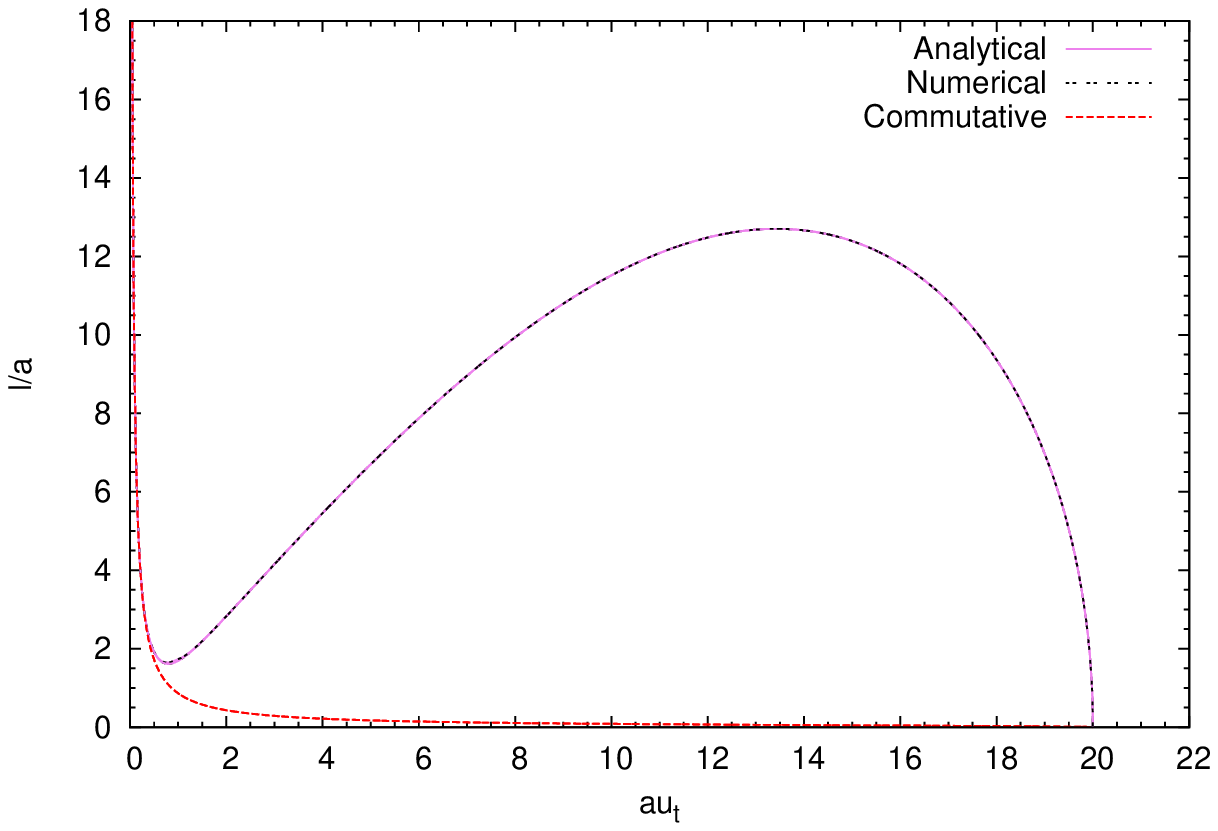}\\
		{\footnotesize $au_b=20$}
	\end{minipage}
	\caption{Variation of $\frac{l}{a}$ with respect to $au_t$ for two different values of the cutoff ($au_{b}=10,20$). The solid curve represents the analytical results given in eqs.(\ref{l2},\ref{L2}) and the dotted curve shows the numerical result. The red dotted curve shows the commutative result. For the analytical curves corresponding to the noncommutative case, the plots have been made using eq.(s)(\eqref{l2},\eqref{L2}). The value of both the functions matches at $au_t=1$ and is equal to $1.7255$ (for $au_b=10$) and $1.733$ (for $au_b=20$).}
	\label{fig1}
\end{figure}\\
\noindent From Fig.(\ref{fig1}), it can be observed that our analytically computed result is in good agreement with the numerically computed result. The plots have been made for two values of cutoff $au_{b}=10,20$. Further, $\frac{l}{a}$ has the first local minimum $\left(\frac{l}{a}\right)_{min} \approx 1.61$ which occurs at $(au_t)_c^{num}\approx 0.78$ (obtained numerically), and $\left(\frac{l}{a}\right)_{min} \approx 1.64$ at $(au_t)_c^{appr}\approx 0.77$ (obtained using eq.\eqref{l2})\footnote{$(au_t)_c$ is the value of $au_t$ where $\frac{l}{a}$ has the first local minimum.}. This in turn means that the domain upto $(au_t)_c$ can be interpreted as the IR domain, and beyond $(au_t)_c$ it probes the deep noncommutative domain (where $\frac{l}{a}$ is proportional to $au_{t}$), and then the deep UV domain. One can also analytically estimate the value of $(au_t)_c$ by using the expressions given in eq.(s)\eqref{lIR},\eqref{lUV}. 
Equating the expressions of $\frac{l}{a}$ corresponding to deep NC and deep IR limits at $au_t = (au_t)_c$, leads to the following
\begin{eqnarray}
\frac{2}{(au_{t})_c^{ana}}\sqrt{\pi}\frac{\Gamma(2/3)}{\Gamma(1/6)} &=& \frac{\sqrt{\pi}}{3}\frac{\Gamma(1/3)}{\Gamma(5/6)}(au_{t})_c^{ana}\nonumber\\
\Rightarrow(au_t)_c^{ana} &=&0.784~.
\end{eqnarray}
The above analytically estimated value of $(au_t)_c^{ana}$ matches well with that obtained graphically (using the approximate expression for $\frac{l}{a}$ given in eq.\eqref{l2}) and numerically. From Fig.(\ref{fig1}), it can be seen that $l$ goes to zero for large $au_{t}$ (that is, $au_{t}\rightarrow au_{b}$) reflecting the fact that extremal surfaces exist for any $l$ .\\ 
Now we shall compute the expression for $a^2S_{EE}$ (given in eq.(\ref{EE1})). Firstly, we compute the expressions corresponding to the deep IR and deep NC limits. As we have observed earlier, in the deep IR limit, the commutative results appear. By using this fact and eq.(\ref{lIR}), we obtain the finite piece of HEE in the deep IR (commutative limit). This reads
\begin{eqnarray}\label{EE4}
\left(a^2\bar{S}_{EE}|^{finite}\right)_{deep~IR}=-\frac{\sqrt{\pi}}{4}\frac{\Gamma(2/3)}{\Gamma(1/6)}(au_{t})^2= 	-(\pi)^{3/2}\left(\frac{\Gamma(2/3)}{\Gamma(1/6)}\right)^3\left(\frac{a}{l}\right)^{2}
\end{eqnarray}
where we have used the scaling $\bar{S}_{EE}= \left(\frac{g_s^2G_N^{(10)}}{R^8 L^2 \mathrm{Vol}(\Omega_{5})}\right)S_{EE}$. On the other hand in the deep NC limit $(\frac{1}{au_{t}}\approx 0)$, the finite piece of HEE reads
\begin{eqnarray}\label{EE3}
\left(a^2\bar{S}_{EE}|^{finite}\right)_{deep~NC} = \frac{1}{16\pi^{3/2}}\left(\frac{3\Gamma(5/6)}{\Gamma(1/3)}\right)^3 \left(\frac{l}{a}\right)^4~.
\end{eqnarray}
Now by following the same procedure we have used to compute $\frac{l}{a}$, the general expression (which can probe the UV/IR mixing) for $a^2\bar{S}_{EE}$ reads (for $au_{t}\le1$)
\begin{eqnarray}\label{EEzero}
a^2\bar{S}_{EE}&=&\frac{(au_{t})^{2}}{2}\bigg[(au_{t})^{2}\int_{\frac{au_{t}}{au_{b}}}^{au_{t}}~dp\frac{\sqrt{1+(\frac{p}{au_{t}})^{4}}}{p^{5}\sqrt{1-p^{6}}}+\int_{au_{t}}^{1}~dp\frac{\sqrt{1+(\frac{au_{t}}{p})^{4}}}{p^{3}\sqrt{1-p^{6}}}\bigg]\label{sint}\nonumber\\
&\approx&a^{2}\bar{S}_{div}+\left(\sum_{n=1}^{\infty}\frac{1}{2\sqrt{\pi}}\frac{\Gamma(n+\frac{1}{2})}{\Gamma(n+1)} \left[\frac{1}{(6n-4)}+\frac{1}{(12n)}-\frac{1}{(6n-2)}\right]\right)(au_t)^{(6n)}\nonumber\\
&-&\sum_{n=2}^{\infty}\frac{1}{4\sqrt{\pi}}\frac{\Gamma(n+\frac{1}{2})}{\Gamma(n+1)}\frac{(au_t)^{(6n)}}{(6n-6)}+\sum_{n=2}^{\infty}\frac{1}{4\sqrt{\pi}}\frac{\Gamma(n+\frac{1}{2})}{\Gamma(n+1)}\frac{(au_t)^{6}}{(6n-6)}+\sum_{n=0}^{\infty}\frac{1}{2\sqrt{\pi}}\frac{\Gamma(n+\frac{1}{2})}{\Gamma(n+1)}\frac{(au_{t})^{2}}{(6n-2)}\nonumber\\
&+&\left[\frac{1}{6}-\frac{(au_{t})^{6}}{24}-(au_{t})^{6}\frac{\Gamma(\frac{3}{2})\log(au_{t})}{4\sqrt{\pi}}\right]~.
\end{eqnarray}
In the above expression, the subsystem information (turning point) independent divergent piece \cite{Srednicki:1993im} reads 
\begin{eqnarray}\label{Sdiv}
a^{2}\bar{S}_{div}= \frac{1}{8}(au_{b})^{4}+\frac{\log(au_{b})}{4}~.
\end{eqnarray}
In the limit $a\rightarrow\frac{1}{u_{b}}$, the finite piece of eq.(\ref{EEzero}) produces the HEE corresponding to the commutative SYM (given in eq.(\ref{EE4}))
\begin{equation}\label{EE2}
a^2\bar{S}_{EE}|_{deep~IR}=\sum_{n=0}^{\infty}\frac{1}{2\sqrt{\pi}}\frac{\Gamma(n+\frac{1}{2})}{\Gamma(n+1)}\frac{1}{(6n-2)}(au_{t})^{2} = -\frac{\sqrt{\pi}}{4}\frac{\Gamma(2/3)}{\Gamma(1/6)}(au_{t})^2~.
\end{equation}
For $au_{t}\ge 1$, the expression for the HEE reads
\begin{eqnarray}\label{see1}
a^{2}\bar{S}_{EE}&\approx&\frac{1}{8}\left((au_{b})^{4}-(au_{t})^{4}\right)-\frac{1}{4}\ln(\frac{au_{t}}{au_{b}})+\frac{(au_{t})^{4}}{2}\sum_{n=1}^{\infty}\frac{1}{\sqrt{\pi}}\frac{\Gamma(n+\frac{1}{2})}{\Gamma(n+1)}\frac{1}{(6n-4)}\left(1-\left(\frac{au_{t}}{au_{b}}\right)^{6n-4}\right)\nonumber\\
&+&\frac{1}{4}\sum_{n=1}^{\infty}\frac{1}{\sqrt{\pi}}\frac{\Gamma(n+\frac{1}{2})}{\Gamma(n+1)}\frac{1}{6n}\left(1-\left(\frac{au_{t}}{au_{b}}\right)^{6n}\right)\nonumber\\
&+&\frac{1}{4}\sum_{m=2}^{\infty}\frac{\sqrt{\pi}}{\Gamma(m+1)\Gamma(\frac{3}{2}-m)}\frac{1}{(au_{t})^{4(m-1)}}\frac{\left(1-\left(\frac{au_{t}}{au_{b}}\right)^{4(m-1)}\right)}{4(m-1)}\nonumber\\
&+&\frac{1}{4}\sum_{n=1}^{\infty}\sum_{m=2}^{\infty}\frac{\Gamma(n+\frac{1}{2})}{\Gamma(n+1)\Gamma(m+1)\Gamma(\frac{3}{2}-m)}
\frac{1}{(au_{t})^{4(m-1)}}\frac{1}{(6n+4(m-1))}\left(1-\left(\frac{au_{t}}{au_{b}}\right)^{6n+4(m-1)}\right)~.\nonumber\\
\end{eqnarray}\\
\begin{figure}[!h]
	\begin{minipage}[t]{0.48\textwidth}
		\centering\includegraphics[width=\textwidth]{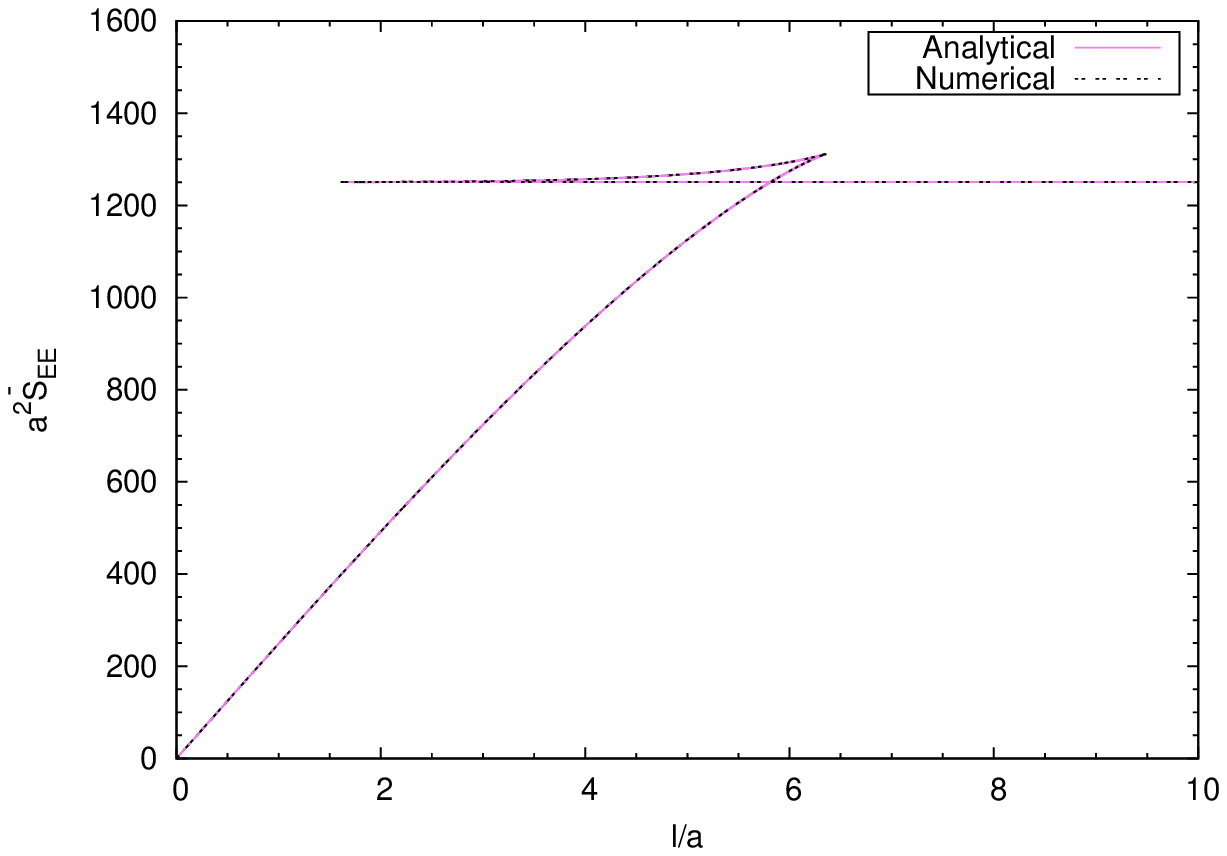}\\
		{\footnotesize $au_b=10$}
	\end{minipage}\hfill
	\begin{minipage}[t]{0.48\textwidth}
		\centering\includegraphics[width=\textwidth]{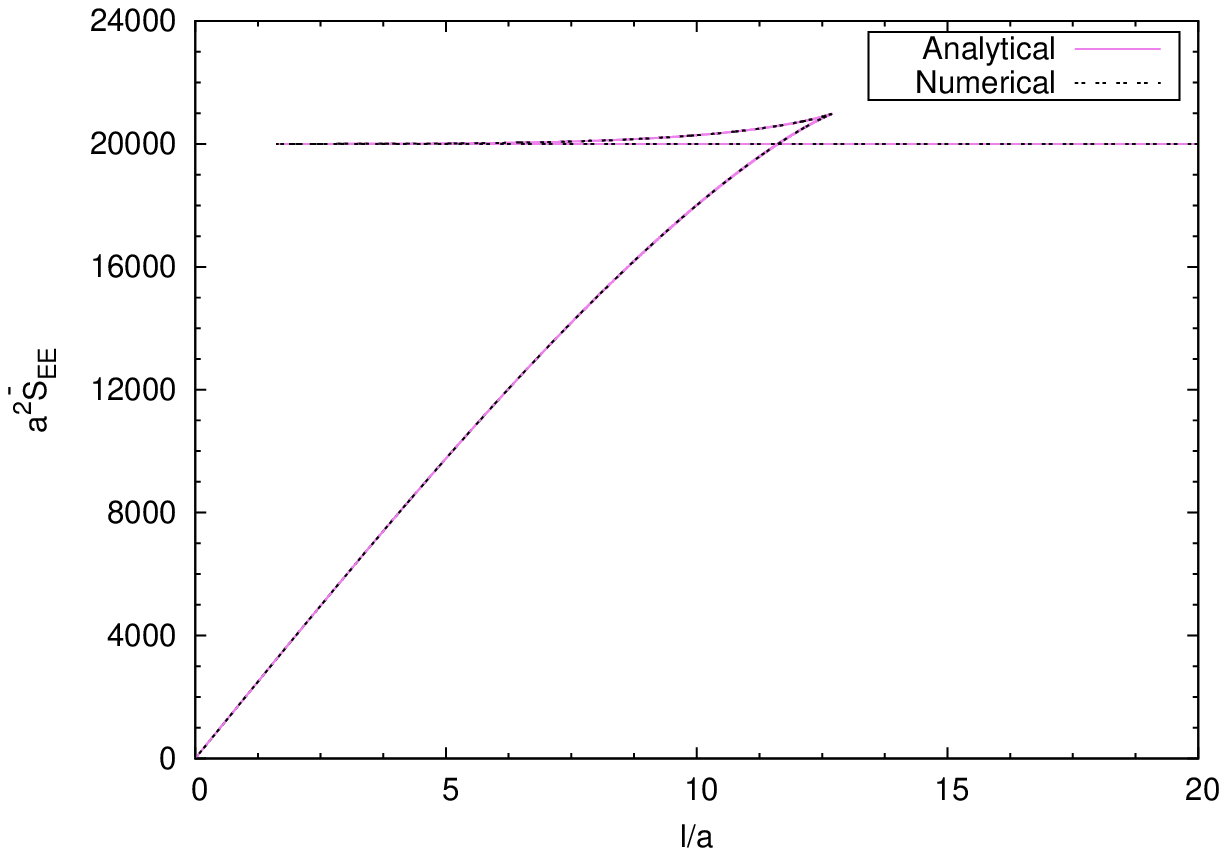}\\
		{\footnotesize $au_b=20$}
	\end{minipage}
	\caption{Variation of $a^2\bar{S}_{EE}$ with respect to $\frac{l}{a}$ for two different values of cutoff ($au_{b}=10,20$). The solid curve represents the analytical results which is obtained by using eq.(s)(\eqref{l2},\eqref{L2},\eqref{EEzero},\eqref{see1}) and the dotted curve represents the numerical result. For the analytical curves corresponding to the noncommutative case, the value of both the functions matches at $au_t=1$ and is equal to $1250$ (for $au_b=10$) and $2\times10^4$ (for $au_b=20$).} 
	\label{fignew}
\end{figure}\\
\noindent It is to be noted that unlike the computed result of $\frac{l}{a}$, the computed result of HEE does not produce the commutative result if we naively take the limit $a\rightarrow0$. The commutative result emerges in the limit $a\rightarrow\frac{1}{u_{b}}$ (which can also be written down as $\sqrt{\vartheta}\rightarrow\frac{1}{\lambda^{1/4}u_b}$). This is due to the reason that we have modified the divergence structure by introducing a dimensionless cut-off $au_b$. This modification relates the NC parameter with the cut-off which is reminiscent of the UV/IR mixing property and the reason is the following. We note that the radial cut-off $u_b$ represents the momentum cut-off of the dual field theory (which is inversely proportional to the lattice spacing). Further, for a noncommutative field theory (with noncommutative parameter $\vartheta$), this momentum cut-off $u_b$ of the lattice field theory is related with the NC parameter $\vartheta$. This in turn means that one cannot take a limiting value of $a=\lambda^{1/4}\sqrt{\vartheta}$ without influencing the momentum cut-off $u_b$. In fig.(\ref{fignew}), we compare our analytically computed result $a^2\bar{S}_{EE}$ (given in eqs.(\ref{EEzero}~,~\ref{see1})) with that obtained numerically from eq.\eqref{EE1}. We have used only the finite pieces of EE. It can be observed that our analytical result is in very good agreement with that obtained numerically. The plots have been made both numerically and analytically for $au_{b}=10,20$.\\
We shall now obtain a critical length $l_{c}$
below which we have the deep UV limit. It has been argued in \cite{Karczmarek:2013xxa} that for studying surfaces anchored on small strips, $u(x_{2})$ has to be expanded in a power series of $x_{2}$ (for small $x_{2}$)
\begin{equation}\label{a}
u(x_{2})=u_{0}+u_{1}x_{2}+u_{2}x_{2}^{2}+...~.
\end{equation}
Substituting this in eq.(\ref{e}) and using $u(x_{2}=0)=u_{t}$  and the boundary condition in eq.(\ref{bc1}), we get 
\begin{equation}
u(x_{2})=u_{t}+\frac{3}{2}\frac{u_{t}^{3}}{[1+a^{4}u_{t}^{4}]}x_{2}^{2}+...~.
\end{equation}
Now putting $x_{2}=\frac{l}{2}$ in the above equation gives \cite{Karczmarek:2013xxa} 
\begin{equation}\label{llc}
u_{b}=u_{t}+\frac{3}{8}\frac{u_{t}^{3}}{[1+(au_{t})^{4}]}l^{2}+\mathcal{O}\left((l/a)^{4}\right)~.
\end{equation}
This result can be substituted in eq.(\ref{RTF}) to get 
\begin{eqnarray}\label{ee2}
a^{2}\bar{S}_{EE}&=&\frac{a^{2}}{2}\bigg(\frac{l}{\epsilon^{3}}-\frac{3}{8}\frac{l^{3}}{\epsilon a^{4}}\frac{1}{[1+(\epsilon/a)^{4}]}-\frac{9}{8}\frac{l^{5}}{a^{8}}\frac{\epsilon}{[1+(\epsilon/a)^{4}]^{2}}+\mathcal{O}\left((l/a)^{7}\right)\bigg)\nonumber\\
&=&\frac{1}{2}\left((au_{b})^{3}\frac{l}{a}-\frac{3}{8}\left(\frac{l}{a}\right)^{3}\frac{(au_{b})}{\left(1+(1/au_{b})^{4}\right)}-\frac{9}{8}\left(\frac{l}{a}\right)^{5}\frac{1}{au_{b}}\frac{1}{\left(1+(1/au_{b})^{4}\right)^{2}}\right)\nonumber\\
&+&\mathcal{O}\left((l/a)^{7}\right)~.
\end{eqnarray}
Note that we have also provided the finite terms which go to zero in the limit $\epsilon\rightarrow 0$ in the above equation. Comparing the leading order divergence term of the above result with that obtained in eq.(\ref{Sdiv}) yields the critical length $l_{c}$ to be \footnote{The left panel of Fig.(\ref{fig1}) has been plotted for $au_{b}=10$. This gives $\frac{l_{c}}{a}=5$. Hence, we can see that the critical length $\frac{l_{c}}{a}$ below which we have the deep UV limit is larger than $\left(\frac{l}{a}\right)_{min}$ which means that an observer in the field theory will not realise that there is a minimum length $\left(\frac{l}{a}\right)_{min}$.} 
\begin{equation}\label{UVL}
l_{c}\approx\frac{a^{2}u_{b}}{2}~.
\end{equation}
Hence, eq.(\ref{ee2}) holds for $l<l_{c}$ and eqs.(\ref{EEzero}~,~\ref{see1}) holds for $l>l_{c}$. Using the relation given in eq.\eqref{UVL}, one can recast the expression of $S_{EE}$ (given in eq.\eqref{ee2}) in the following form
\begin{eqnarray}\label{EEUV}
a^{2}\bar{S}_{EE}&=&\frac{a^{2}}{2}\left(\frac{ l_cl}{a}\right)\bigg[\frac{8l_c^2}{a^5}-\frac{3}{4}\left(\frac{1}{1+(\frac{a}{2l_c})^{4}}\right)\left(\frac{l^2}{a^5}\right)-\frac{9}{16}\left(\frac{1}{1+(\frac{a}{2l_c})^{4}}\right)^2\left(\frac{l^4}{l_c^2a^5}\right)+....\bigg]\nonumber\\
&=&\frac{l_{c}l}{2}\left[\frac{8l_{c}}{a^{3}}-\frac{3}{4}\left(\frac{1}{1+(a/2l_{c})^{4}}\right)\left(\frac{l^{2}}{a^{3}}\right)-\frac{9}{16}\left(\frac{1}{1+(a/2l_{c})^{4}}\right)^{2}\left(\frac{l^{4}}{a^{3}l_{c}^{2}}\right)+...\right]~.
\end{eqnarray}
We shall now investigate the $c$-function of the dual field theory holographically. We shall carry out our investigation for $l>l_{c}$.
As we shall see in the subsequent discussion that the deep UV solution (for $l<l_{c}$) poses problems in the determination of the $c$-function.
For this we now proceed to write down the expression of $a^2\bar{S}_{EE}|^{finite}$ (given in eq.(\ref{EEzero}) for $au_{t}\le 1$) in terms of the dimensionless form of the subsystem size $\frac{l}{a}$ (given in eq.(\ref{l2}) for $au_{t}\le1$). Keeping the leading order noncommutative correction, eq.(\ref{l2}) can be recast as
\begin{equation}
\frac{l}{a}\simeq\frac{\sqrt{\pi}}{2(au_{t})}\frac{\Gamma(\frac{5}{3})}{\Gamma(\frac{7}{6})}\left[1+\frac{1}{3\sqrt{\pi}}\frac{\Gamma(7/6)}{\Gamma(5/3)}\left(4+\log4-6\log(au_t)-6\left(\frac{1}{au_b}\right)^2-\left(\frac{1}{au_b}\right)^6\right)(au_t)^4\right]~~.
\end{equation}
Now with the above expression in hand and assuming $au_t$ to be very small ($au_{t}<<1$), we can solve it perturbatively and write down $au_t$ in terms of $\frac{l}{a}$. This reads (for $au_{t}<<1$)
\begin{equation}\label{tpoint1}
au_{t}\left(\frac{l}{a}\right)=\frac{\alpha_0}{(\frac{l}{a})}+\frac{\alpha_1}{(\frac{l}{a})^5}+\frac{\alpha_0^4\log(\frac{l}{a})}{(\frac{l}{a})^5}~
\end{equation}
where $\frac{au_{t}}{au_{b}}$ has been neglected since it is very small.
Similarly, we now approximate the expression for $a^2\bar{S}_{EE}|^{finite}$ (given in eq.\eqref{EEzero}) by keeping the leading order NC correction terms only. This leads to the following expression (for $au_{t}<< 1$)
\begin{eqnarray}\label{EEapprox}
a^2\bar{S}_{EE}|^{finite} \simeq \frac{1}{6}-\frac{\sqrt{\pi}}{4}\frac{\Gamma(2/3)}{\Gamma(1/6)}(au_t)^2+\frac{1}{48}\left(3+\log 4-6\log(au_t)\right)(au_t)^6~.
\end{eqnarray}
By substituting the expression of turning point $au_t$ (given in eq.\eqref{tpoint1}) in eq.\eqref{EEapprox}, we obtain
\begin{eqnarray}\label{EEl}
a^2\bar{S}_{EE}\left(\frac{l}{a}\right)|^{finite}&=&\frac{1}{6}-\left[\frac{\sqrt{\pi}}{4}\frac{\Gamma(2/3)}{\Gamma(1/6)}-\frac{(3+\log 4)}{48}\left[\frac{\alpha_0}{(\frac{l}{a})}+\frac{\alpha_1}{(\frac{l}{a})^5}+\frac{\alpha_0^4\log(\frac{l}{a})}{(\frac{l}{a})^5}\right]^4\right]\nonumber\\
&\times&\left[\frac{\alpha_0}{(\frac{l}{a})}+\frac{\alpha_1}{(\frac{l}{a})^5}+\frac{\alpha_0^4\log(\frac{l}{a})}{(\frac{l}{a})^5}\right]^2-\left(\frac{1}{8}\right)\left[\frac{\alpha_0}{(\frac{l}{a})}+\frac{\alpha_1}{(\frac{l}{a})^5}+\frac{\alpha_0^4\log(\frac{l}{a})}{(\frac{l}{a})^5}\right]^6\nonumber\\
&\times&\log\left[\frac{\alpha_0}{(\frac{l}{a})}+\frac{\alpha_1}{(\frac{l}{a})^5}+\frac{\alpha_0^4\log(\frac{l}{a})}{(\frac{l}{a})^5}\right]
\end{eqnarray}
where
\begin{eqnarray}
\alpha_0= \frac{\sqrt{\pi}}{2}\frac{\Gamma(\frac{5}{3})}{\Gamma(\frac{7}{6})};~~ ~\alpha_1=\frac{2}{3}\alpha_0^4+\frac{\alpha_0^4}{6}\log 4-\alpha_0^4\log\alpha_0\nonumber~.
\end{eqnarray}
The expression given in eq.\eqref{EEl} represents HEE for a strip-like subsystem at the boundary. Using the above expression we can holographically compute the $c$-function of the  dual field theory which we shall carry out in the next section.
\section{Holographic computation of the c-function}\label{newsection}
\noindent
The $c$-function is a monotonically decreasing function (under renormalization group flow) measuring the degrees of freedom of the theory and is stationary at the fixed points of the renormalization group flow. Further, the value of the $c$-function at the fixed points are related to the central charge of the two-dimensional conformal field theory (CFT). 
In \cite{Casini:2004bw,Casini:2006es}, a $c$-function in terms of the entanglement entropy was computed for two-dimensional CFT. This was a entropic reformulation of the Zamolodchikov theorem \cite{Zamolodchikov:1986gt}. 

\noindent For the EE corresponding to a single interval of length $l$, the $c$-function for 2D CFT reads \cite{Casini:2004bw,Casini:2006es,Klebanov:2007ws}
\begin{eqnarray}
c&=&  3l\frac{dS_{EE}}{dl}~.
\end{eqnarray}\\
Following this direction, in \cite{Myers:2012ed} a $c$-function in terms of the EE, for a $d+1$-dimensional CFT has been proposed. It is known that the HEE corresponding to a `slab' like subsystem is given by \cite{Ryu:2006ef}
\begin{eqnarray}\label{Ee}
S_{EE} &=& \alpha \frac{L^{d-1}}{\epsilon^{d-1}} - \frac{1}{(d-1)} \frac{C_d}{\beta} \left(\frac{L}{l}\right)^{d-1}
\end{eqnarray}
where $\alpha$ and $\beta$ are dimensionless constants, $\epsilon$ is the UV regulator and $C_d$ is the central charge. Following the idea of 2D CFT, the following $c$-function along the RG flow for a $d+1$-dimensional CFT was proposed \cite{Myers:2012ed}
\begin{eqnarray}\label{CF}
c=\left(\frac{\beta}{L^{d-1}}\right)l^d\frac{dS_{EE}}{dl}~.
\end{eqnarray}
Without loss of generality, we use the above $c$-function to characterize the degrees of freedom of NC SYM. It is to be noted that the above mentioned $c$-function has been proposed for Lorentz invariant theories. On the other hand the full Lorentz symmetry for NC SYM is broken as $SO(3,1)\rightarrow SO(1,1) \times SO(2)$. It remains an open problem to construct a $c$-function for systems with broken Lorentz symmetry. However, we shall use the above definition of the $c$-function since it can still probe the degrees of freedom of the system and observe the effect of noncommutativity on it. Firstly, we look at the deep IR limit of the NC SYM. In this limit, we have the commutative SYM and the EE reads (given in \eqref{EE4})
\begin{eqnarray}
S_{EE}=\left(\frac{R^8L^2 \mathrm{Vol}(\Omega_5)}{g_s^2 G_N^{(10)}}\right)\left[S_{div}^{SYM}	-\pi^{3/2}\left(\frac{\Gamma(2/3)}{\Gamma(1/6)}\right)^3\left(\frac{1}{l}\right)^{2}\right]
\end{eqnarray}
where $S_{div}^{SYM}$ represents the universal divergent term of SYM. Now by correctly identifying $\beta$ and by using the definition (given in \eqref{CF}) for $d=3$, the $c$-function for the commutative SYM is obtained to be
\begin{eqnarray}
c=\frac{2R^8 \mathrm{Vol}(\Omega_5)}{g_s^2 G_N^{(10)}}=C^{sym}~.
\end{eqnarray}
This is also the central charge of the $\mathcal{N}=4$ SYM theory in $3+1$-dimensions. It is to be noted that this identification of dimensionless quantity $\beta$ is difficult to carry out for the expression of $S_{EE}$ corresponding to NC SYM (given in \eqref{EEzero}). Hence, we proceed with the expression given in eq.(\ref{EEl}). We can recast the expression in the following form
\begin{eqnarray}
S_{EE}&=&-\frac{C^{sym}L^2}{2}\frac{1}{\beta_{0}}\left(\frac{1}{l}\right)^2+C^{sym}L^2\left[\frac{1}{12a^2}-\frac{a^4\alpha_{0}^6}{16l^6}\log(\frac{\alpha_{0}a}{l})\right]\nonumber\\
&+&C^{sym}L^2\frac{a^4\alpha_{0}}{l^6}\left[\frac{\alpha_{0}^5(3+\log 4)}{96}-\frac{\sqrt{\pi}}{4}\frac{\Gamma(2/3)}{\Gamma(1/6)}\alpha_{1}-\frac{\sqrt{\pi}}{4}\frac{\Gamma(2/3)}{\Gamma(1/6)}\alpha_{0}^4\log(\frac{l}{a})\right]
\end{eqnarray}
where we have identified $C^{sym}=\frac{R^{8}Vol(\Omega_{5})}{g_{s}^{2}G_{N}^{10}}$ and $\beta_{0}=\frac{4\Gamma(1/6)}{\sqrt{\pi}\Gamma(2/3)\alpha_{0}^{2}}$. It is to be noted that the first term is the usual one which we get from the commutative SYM in $3+1$-spacetime dimensions. We now introduce a $l$-dependent $\beta$, namely, $\beta(l)$ and recast the above expression in the form
\begin{equation}\label{Cnew}
S_{EE}=\frac{C^{sym}L^2}{12a^2}-\frac{C^{sym}L^2}{2\beta(l)l^2}
\end{equation}
where
\begin{eqnarray}
\frac{1}{\beta(l)}&=&\frac{1}{\beta_{0}}-\alpha_{0}\left(\frac{a}{l}\right)^4\bigg\{\frac{\alpha_{0}^5(3+\log 4)}{48}-\frac{\sqrt{\pi}}{2}\frac{\Gamma(2/3)}{\Gamma(1/6)}\alpha_{1}-\frac{\sqrt{\pi}}{2}\frac{\Gamma(2/3)}{\Gamma(1/6)}\alpha_{0}^4\log(\frac{l}{a})\nonumber\\
&-&\frac{\alpha_{0}^5}{8}\log(\frac{\alpha_{0}a}{l})\bigg\}
\end{eqnarray}
\noindent We now define the entropic $c$-function in the following way
\begin{eqnarray}
c&=&\frac{\beta(l)}{L^{2}}l^3\frac{dS_{EE}}{dl}~.
\end{eqnarray}
Computation of the above expression leads to the following
\begin{eqnarray}\label{cf}
\frac{c}{C^{sym}}&=& 1-\beta_{0}\left(\frac{a}{l}\right)^4\alpha_{0}\Bigg[\frac{(3+\log 4)}{24}\alpha_0^5-\sqrt{\pi}\frac{\Gamma(2/3)}{\Gamma(1/6)}\alpha_{1}-\sqrt{\pi}\frac{\Gamma(2/3)}{\Gamma(1/6)}\alpha_{0}^4\log(\frac{l}{a})-\frac{\alpha_{0}^5}{4}\log(\frac{\alpha_{0}a}{l})\nonumber\\
&+&\frac{\sqrt{\pi}}{4}\alpha_0^4\frac{\Gamma(2/3)}{\Gamma(1/6)}+\frac{\alpha_{0}^5}{16}\Bigg]~.
\end{eqnarray}
\noindent It can be observed that the first term in the above equation is the central charge of the commutative theory whereas the rest probes the signature of a Lorentz violating theory induced by noncommutativity. 
It can be noted that in the deep IR limit the $c$-function of the NCYM approaches the constant value $C^{sym}$ corresponding to the commutative Yang-Mills theory. For large $\frac{l}{a}$, (for $au_{b}=10$), eq.(\ref{cf}) yields $\frac{c}{C^{sym}}=1$ which agrees very well with the numerical result. There is a small difference between the two results only at the third decimal place.\\
\noindent Before ending this discussion we would like to point out that in the deep UV limit $(l<l_{c})$, the definition of c-function (given in eq.\eqref{CF}) runs into a problem because of the non-locality of the theory which leads to the violation of area law for $S_{EE}$. To see this we compute and graphically represent the $c$-function for all possible values of $\frac{l}{a}$.
\begin{figure}[!h]
	\begin{minipage}[t]{0.48\textwidth}
		\centering\includegraphics[width=\textwidth]{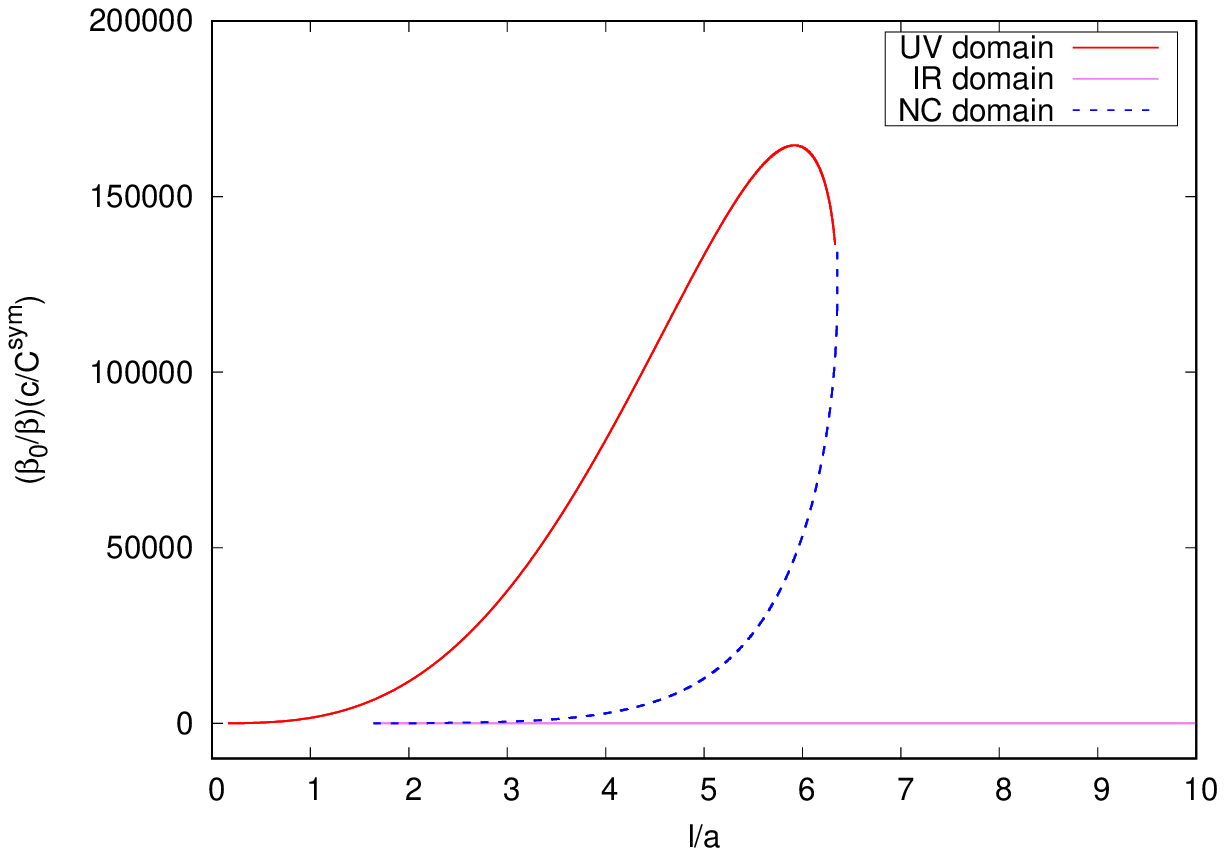}\\
		{\footnotesize Behaviour of the $c$-function in each domain}
	\end{minipage}\hfill
	\begin{minipage}[t]{0.48\textwidth}
		\centering\includegraphics[width=\textwidth]{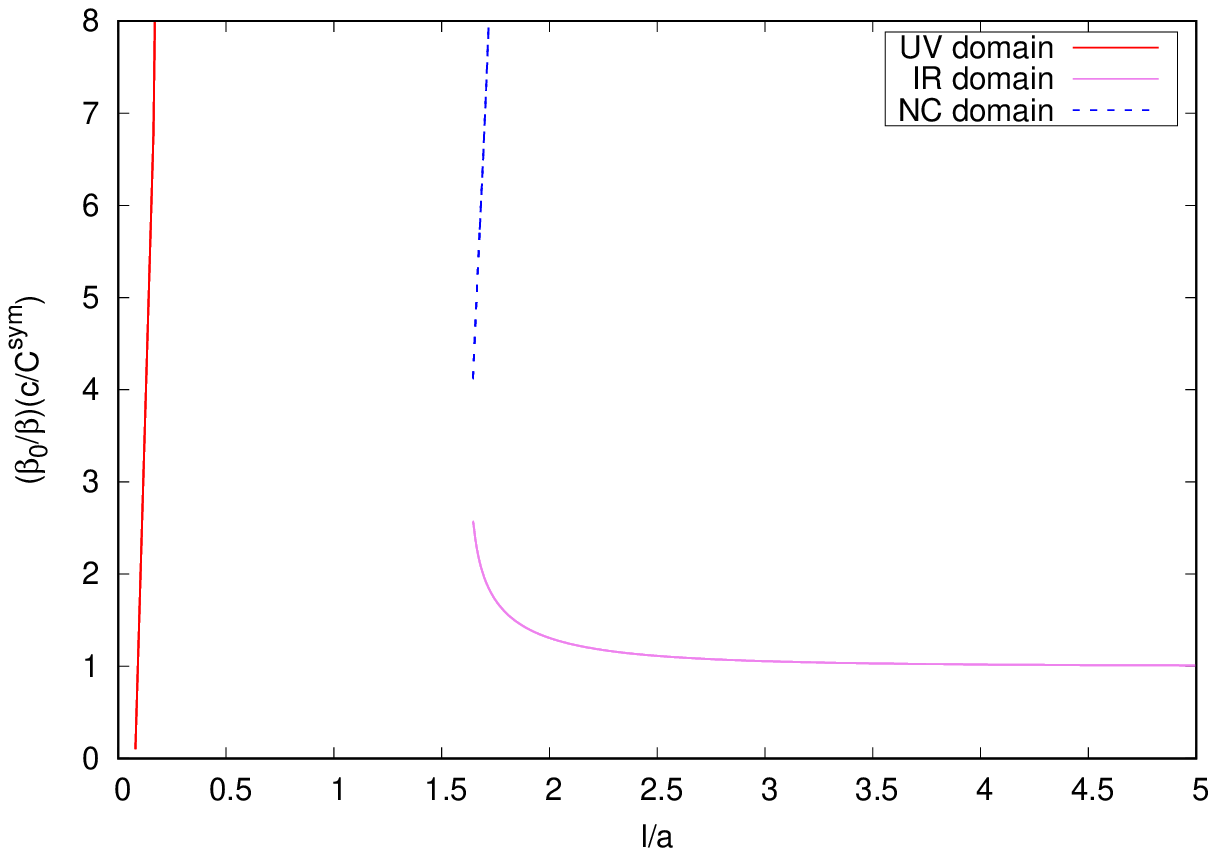}\\
		{\footnotesize Discontinous nature of the $c$-function}
	\end{minipage}\hfill
	\caption{Behaviour of the $c$-function for all possible values of $\frac{l}{a}$ (we set $au_b$=10).}
	\label{figCfn}
\end{figure}\\
In Fig.\eqref{figCfn},  we have plotted $\left(\frac{\beta_{0}}{\beta}\right)\left(\frac{c}{C^{sym}}\right)$ in the vertical axis and in the horizontal axis we have plotted $\frac{l}{a}$. 
In the left panel of Fig.\eqref{figCfn},	we observe that in the IR domain the ratio $\left(\frac{\beta_0}{\beta}\right)\left(\frac{c}{C^{sym}}\right)$ apporaches unity, that is, the $c$-function of NC SYM matches with that of SYM. Interestingly, we observe that there are discontinuous jumps in the quantity (see right panel of Fig.\eqref{figCfn}) which is due to the swallowtail behaviour of the HEE. These jumps appear at the junction between the IR and NC domains of the theory, and the junction between the NC and UV domains (see Fig.(\ref{disc})) of the theory. The discontinuities in the $c$-function therefore correspond to the transitions from one domain to the other, in particular the discontinuity between the IR and the NC domains correspond to a transition from the area to the volume law for the HEE.
\begin{figure}
	\centering
	\includegraphics[width=0.5\textwidth]{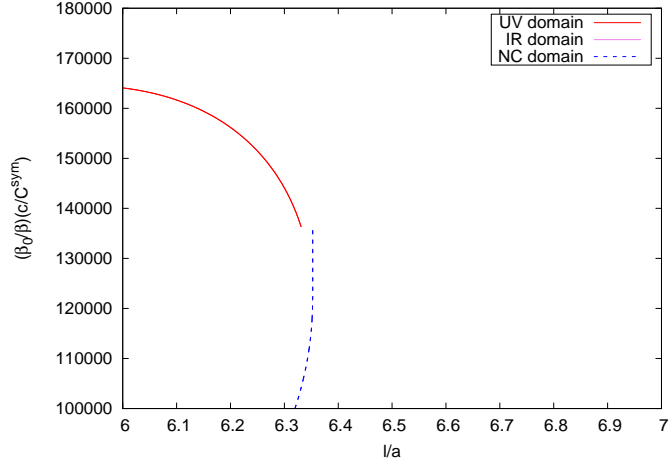}
	\caption{Discontinuty in the $c$-function at the UV and the NC junction.}
	\label{disc}
\end{figure}
\section{Entanglement wedge cross section}\label{Sec3}
In this section we compute the EWCS for NC Yang-Mills gauge theory. This computation holographically probes the entanglement of purification on the basis of $E_P = E_W$ duality \cite{Takayanagi:2017knl}. We proceed by considering two strip-like subsystems on the boundary $\partial M$ ($\partial M$ is the boundary of a time-slice $M$ we have considered in the gravity dual). We denote these subsystems as $A$ and $B$ with both of them having the same length $l$. Further we consider that $A$ and $B$ are separated by a distance $d$ with the condition $A\cap B =0$. The Ryu-Takayanagi surfaces corresponding to $A$, $B$ and $AB$ are denoted as $\Gamma_A^{min}$, $\Gamma_B^{min}$ and $\Gamma_{AB}^{min}$ respectively. The codimension-0 domain of entanglement wedge $M_{AB}$ is characterized by the following boundary 
\begin{eqnarray}\label{16}
\partial M_{AB} = A \cup B \cup \Gamma_{AB}^{min} =\bar{\Gamma}_A \cup \bar{\Gamma}_B
\end{eqnarray}
where $\bar{\Gamma}_A = A \cup \Gamma_{AB}^{A}$, $\bar{\Gamma}_B = B \cup \Gamma_{AB}^{B}$. In the above equation we have used the condition $\Gamma_{AB}^{min} = \Gamma_{AB}^{A} \cup \Gamma_{AB}^{B}$. In this set up, one can define the holographic entanglement entropies $S(\rho_{A \cup \Gamma_{AB}^{A}})$ and $S(\rho_{B \cup \Gamma_{AB}^{B}})$ and compute them by finding a static RT surface $\Sigma^{min}_{AB}$ with the following condition
\begin{eqnarray}
\partial \Sigma^{min}_{AB} = \partial \bar{\Gamma}_A =  \partial \bar{\Gamma}_B~.
\end{eqnarray}
The spliting condition $\Gamma_{AB}^{min} = \Gamma_{AB}^{A} \cup \Gamma_{AB}^{B}$ which has been incorporated in not unique and there can be infinite number of possible choices. Further, this means that there can be infinite number of choices for the surface $\Sigma^{min}_{AB}$. The EWCS is computed by minimizing the area of $\Sigma^{min}_{AB}$ over all possible choices for $\Sigma^{min}_{AB}$. This reads
\begin{eqnarray}\label{18}
E_W(\rho_{AB}) = \mathop{min}_{\bar{\Gamma}_A \subset \partial M_{AB}}\left[\frac{A\left(\Sigma^{min}_{AB}\right)}{4G_{d+1}}\right]~.
\end{eqnarray}
This in turn means that EWCS is the vertical constant $x_2$ hypersurface with minimal area which splits $M_{AB}$ into two domains corresponding to $A$ and $B$. The time induced metric on this constant $x_2$ hypersurface reads
\begin{eqnarray}
ds_{ind}^2&=&R^{2}\bigg[u^{2}dx_{1}^{2}+u^{2}h(u)dx_{3}^{2}+\frac{du^{2}}{u^{2}}\bigg]+R^{2}d\Omega_{5}^{2}~.
\end{eqnarray}
By using this above mentioned induced metric and the formula given in eq.\eqref{18}, the EWCS is found to be
\begin{eqnarray}\label{Ew1}
a^{2}E_{W}& =&a^{2}\frac{R^8 L^2 \mathrm{Vol}(\Omega_{5})}{4g_s^2G_N^{(10)}}\int_{u_t(2l+d)}^{u_t(d)}u\sqrt{1+a^4u^4} ~du\nonumber\\
&=&\frac{R^8 L^2 \mathrm{Vol}(\Omega_{5})}{4g_s^2G_N^{(10)}}\left[\frac{1}{4}\left((au_{t}(d))^{2}\sqrt{1+(au_{t}(d))^{4}}-(au_{t}(2l+d))^{2}\sqrt{1+(au_{t}(2l+d))^{4}}\right)\right.\nonumber\\
&+&\left.\frac{1}{4}\left(\sinh[-1]((au_{t}(d))^{2})-\sinh[-1]((au_{t}(2l+d))^{2})\right) \right]~~.\label{ewcs}
\end{eqnarray}
It is to be noted that in general the above expression of EWCS always satisfies the bound
\begin{eqnarray}\label{bound}
E_W \geq \frac{1}{2}I(A:B)~.
\end{eqnarray}
We shall check this explicitly in our study.  It has been pointed out in  \cite{Takayanagi:2017knl} that for a given $l$, there exists a critical separation length ($d_c<l$) between the two subsystems $A$ and $B$ above which there is no connected phase. We shall also see this feature in our study. This means that the codimension-0 bulk region $M_{AB}$ (entanglement wedge) will be disconnected and therefore results in vanishing $E_W(\rho_{AB})$. Up to this critical separation length $d_c$, the mutual information $I(A:B)$ is non-zero and the RT surface $\Gamma_{AB}^{min}$ is in connected phase which leads to a non-vanishing $E_W$. However beyond this critical separation length $d_c$, the mutual information $I(A:B)=0$ and $\Gamma_{AB}^{min}$ is in disconnected phase which results in a vanishing $E_W$. The value of this critical separation length $d_c$ can be computed from the vanishing condition of the mutual information at $d=d_c$. This can be formally written as \cite{Fischler:2013gsa,Ben-Ami:2014gsa}
\begin{eqnarray}
I(A:B)=2S_{EE}(l)-S_{EE}(d)-S_{EE}(2l+d)=0~.
\end{eqnarray}
In the above expression we have used the fact $S_{EE}(A\cup B)=S_{EE}(2l+d)+S_{EE}(d)$, for ``small''  $d/l$.\\
For $au_{t}\le 1$, eq.(\ref{ewcs}) can be simplified as
\begin{eqnarray}\label{EwT01}
a^2\bar{E}_W =  \frac{1}{8}\left[\left((au_t\left(d\right)\right)^2-\left(au_t\left(2l+d\right)\right)^2\right]+\frac{1}{32}\left[\left(au_t\left(d\right)\right)^6-\left(au_t\left(2l+d\right)\right)^6\right]
\end{eqnarray}
where $\bar{E}_W=\left(\frac{g_s^2 G_N^{(10)}}{R^8 L^2 \mathrm{Vol}(\Omega_{5})}\right)E_W$. In the above expression, $au_t\left(d\right)$ and $au_t\left(2l+d\right)$ represents the turning points associated with the RT surfaces $\Gamma_{d}^{min}$ and $\Gamma_{2l+d}^{min}$.\\ 
In order to probe the bound given in eq.\eqref{bound}, we need to compute the expression of $I(A:B)$. We do this by using the expression of $a^2\bar{S}_{EE}\left(\frac{l}{a}\right)$ for $au_{t}\le 1$ given in eq.\eqref{EEzero} and the expression for $\frac{l}{a}$ for $au_{t}\le1$ given in eq.(\ref{l2}). The finite piece of the expression in eq.(\ref{EEzero}) contributes in the mutual information, and similarly expressions of HEE for subsystems $(2l+d)$ and $d$ can be obtained from eq.(\ref{EEzero}) which contribute to the mutual information.
\footnote{The divergent pieces of the HEEs are independent of subsystem size information so they cancel out and do not contribute.} This in turn means that we obtain the following dimensionless form of holographic mutual information
\begin{eqnarray}\label{HMI}
a^2\bar{I}(A:B) = 2a^2\bar{S}_{EE}\left(\frac{l}{a}\right)-a^2\bar{S}_{EE}\left(\frac{d}{a}\right)-a^2\bar{S}_{EE}\left(\frac{2l+d}{a}\right)
\end{eqnarray}
where $a^{2}\bar{S}_{EE}$ is given in eq.(\ref{EEzero}), and we have used the scaling $\bar{I}=\left(\frac{g_s^2G_N^{(10)}}{R^8 L^2 \mathrm{Vol}(\Omega_{5})}\right)I$.
Further, in order probe the effect of noncommutativity, we also compute the expression for $E_W$ and $I(A:B)$ in the deep IR (commutative) and deep noncommutative limits.\\
In the deep IR (commutative) limit, the expressions for $E_W$ and $I(A:B)$ (in dimensionless form) reads
\begin{eqnarray}
a^2\bar{E}_W|_{deep~IR} &=& \frac{1}{8} \left(2\sqrt{\pi}\frac{\Gamma(2/3)}{\Gamma(1/6)}\right)^2\left[\frac{1}{\left(\frac{d}{a}\right)^2}-\frac{1}{\left(\frac{2l+d}{a}\right)^2}\right]\label{EwIR}\\
a^2\bar{I}(A:B)|_{deep~IR} &=& -{\pi}^{3/2}\left(\frac{\Gamma(2/3)}{\Gamma(1/6)}\right)^3\left[\frac{2}{\left(\frac{l}{a}\right)^2}-\frac{1}{\left(\frac{d}{a}\right)^2}-\frac{1}{\left(\frac{2l+d}{a}\right)^2}\right]~.\label{HMiIR}
\end{eqnarray}
On the other hand, in the deep noncommutative limit, the expressions for $E_W$ and $I(A:B)$ (in dimensionless form) reads
\begin{eqnarray}
a^2\bar{E}_W|_{deep~NC} &=& \frac{1}{16} \left(\frac{3}{\sqrt{\pi}}\frac{\Gamma(5/6)}{\Gamma(1/3)}\right)^4\left[\left(\frac{d}{a}\right)^4-\left(\frac{2l+d}{a}\right)^4\right]\label{EwUV}\\
a^2\bar{I}(A:B)|_{deep~NC} &=&\frac{1}{16{\pi}^{3/2}}\left(\frac{3\Gamma(5/6)}{\Gamma(1/3)}\right)^3\left[2\left(\frac{l}{a}\right)^4-\left(\frac{d}{a}\right)^4-\left(\frac{2l+d}{a}\right)^4\right]~.
\label{HMiUV}
\end{eqnarray}
In the deep UV limit, we substitute eq.(\ref{llc}) in eq.(\ref{ewcs}) to get
\begin{eqnarray}\label{ewcs4}
16	a^{2}\bar{E}_{W}&=&\left(au_{b}-\frac{3}{8}\frac{(au_{b})^{3}}{1+(au_{b})^{4}}\left(\frac{d}{a}\right)^{2}\right)^{2}\sqrt{1+\left(au_{b}-\frac{3}{8}\frac{(au_{b})^{3}}{1+(au_{b})^{4}}\left(\frac{d}{a}\right)^{2}\right)^{4}}\nonumber\\
&-&\left(au_{b}-\frac{3}{8}\frac{(au_{b})^{3}}{1+(au_{b})^{4}}\left(\frac{2l+d}{a}\right)^{2}\right)^{2}\sqrt{1+\left(au_{b}-\frac{3}{8}\frac{(au_{b})^{3}}{1+(au_{b})^{4}}\left(\frac{2l+d}{a}\right)^{2}\right)^{4}}\nonumber\\
&+&\sinh[-1](\left(au_{b}-\frac{3}{8}\frac{(au_{b})^{3}}{1+(au_{b})^{4}}\left(\frac{d}{a}\right)^{2}\right)^{2})-\sinh[-1](\left(au_{b}-\frac{3}{8}\frac{(au_{b})^{3}}{1+(au_{b})^{4}}\left(\frac{2l+d}{a}\right)^{2}\right)^{2}).~~~~~
\end{eqnarray}
We now graphically represent our results and observe the effect of noncommutativity on the EWCS and holographic mutual information.
\begin{figure}[!h]
	\begin{minipage}[t]{0.48\textwidth}
		\centering\includegraphics[width=\textwidth]{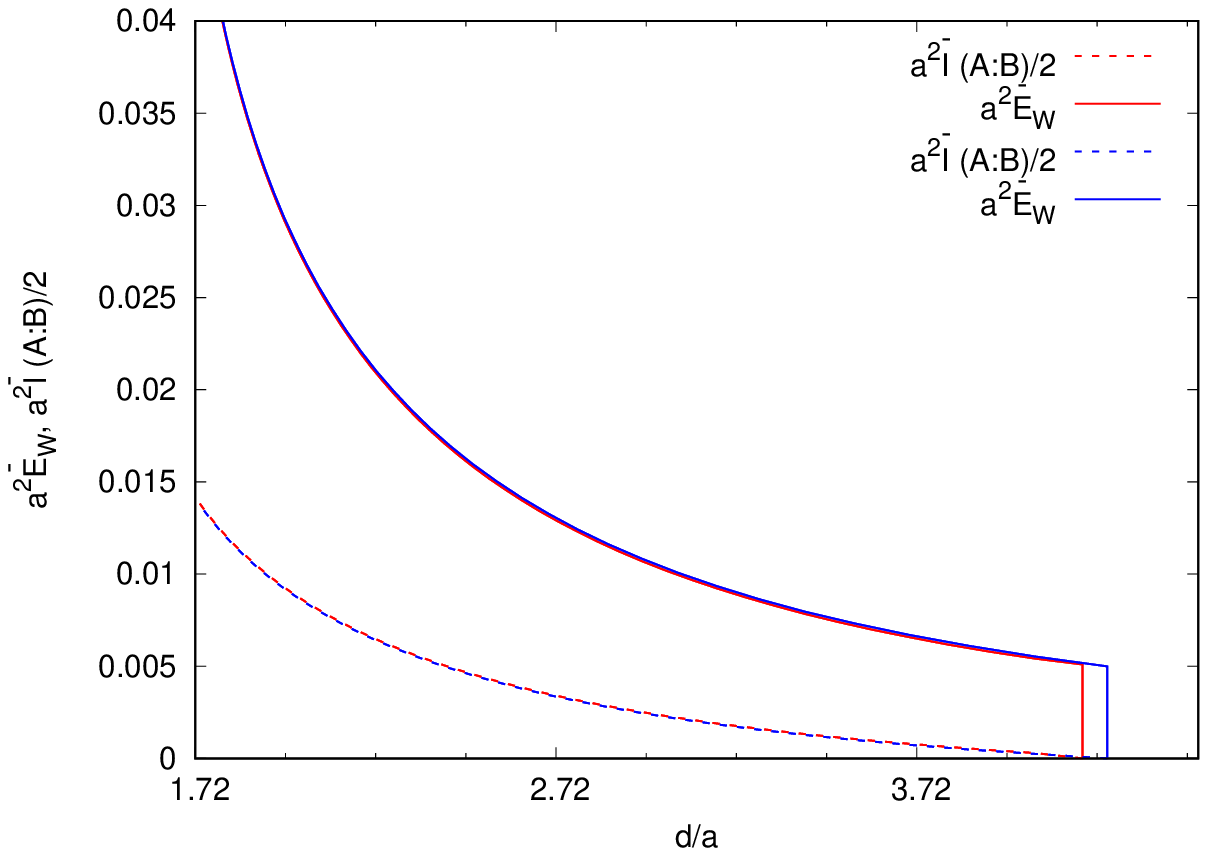}\\
		{\footnotesize Noncommutative Yang-Mills theory}
	\end{minipage}\hfill
	\begin{minipage}[t]{0.48\textwidth}
		\centering\includegraphics[width=\textwidth]{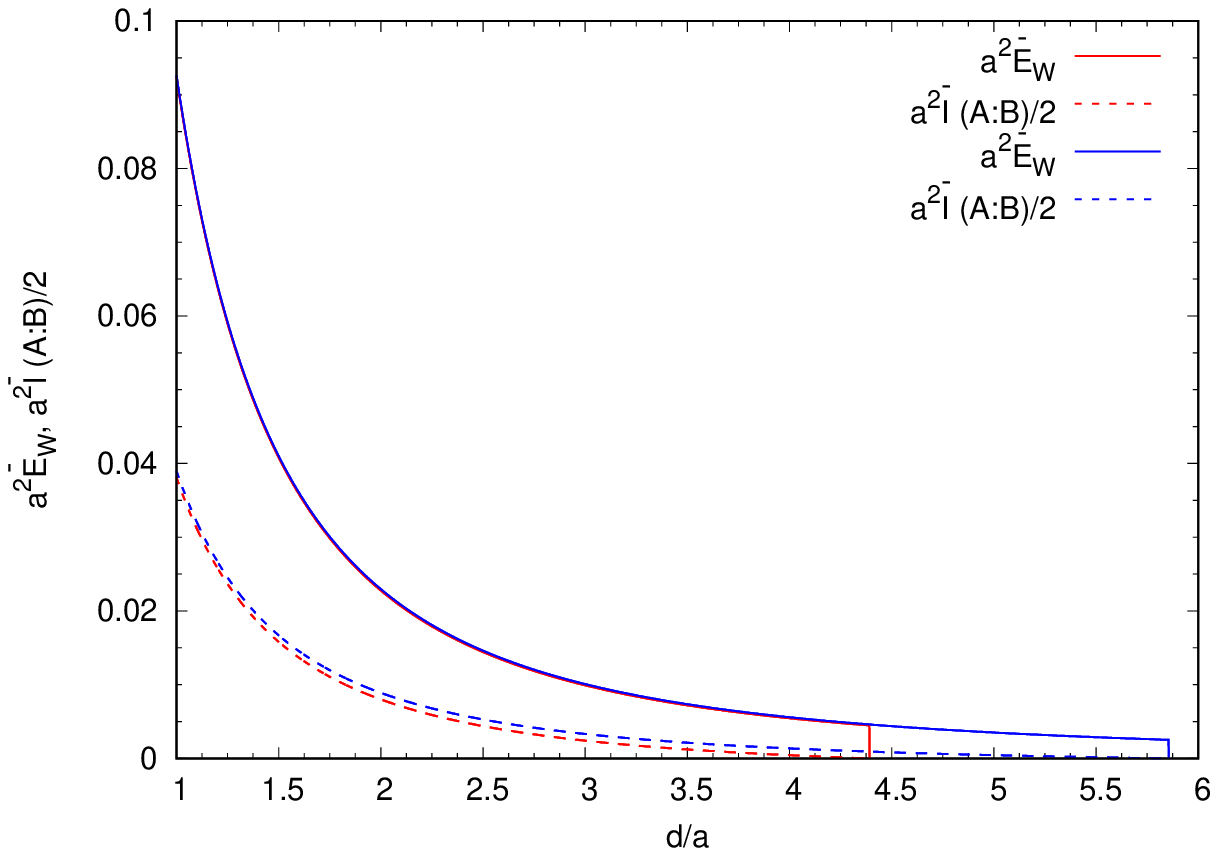}\\
		{\footnotesize Commutative (deep IR) Yang-Mills theory}
	\end{minipage}
	\caption{Effect of noncommutativity on EWCS and HMI (we have chosen two values for $\frac{l}{a}$,  $\frac{l}{a}=6$ (red) and $\frac{l}{a}=8$ (blue)). The curve in the left panel depicts the analytical result in eq.(\ref{EwT01},\ref{HMI}) (for $au_{t}\le 1$) with $au_{b}=10$. The curve in the right panel depicts the analytical results in eqs.(\ref{EwIR},\ref{HMiIR}) (deep IR limit)~. 
	} 
	\label{fig2}
\end{figure}\\
In Fig.\eqref{fig2}, we have plotted $a^2\bar{E}_W$ and $\frac{a^{2}}{2}\bar{I}(A:B)$ with respect to the separation distance $\frac{d}{a}$ for both noncommutative (given in eq.(s)\eqref{EwT01},\eqref{HMI}) and commutative Yang-Mills theory. In getting the plots for $a^{2}\bar{E}_{W}$ and $\frac{a^{2}}{2}\bar{I}(A:B)$, we need eq.(s)(\ref{l2}, \ref{EEzero}) which are valid for $au_{t}\le 1$. The commutative results (given in eq.(s)\eqref{EwIR},\eqref{HMiIR}) are obtained by taking the deep IR limit. We observe that, in both of the above cases, the bound $a^2E_W\geq \frac{1}{2}a^2I(A:B)$ holds and the value of critical separation point or the point of phase transition (from connected to disconnected phase of $M_{AB}$) increases for increase in the value of $\frac{l}{a}$. It is to be noted that for the noncommutative case (in Fig.\eqref{fig2}), we have chosen the lower limit of the $\frac{d}{a}$ axis in such a way so that $au_t(d)\leq1$. Furthermore, the values of $\frac{l}{a}$ are also chosen in order to make $au_t(2l+d),~au_t(l)\leq1$. The domain for which $\frac{d}{a}<1.61$ is not allowed as it corresponds to $au_t>>1$ for which the above analysis breaks down. 
\noindent Further, we note that in the deep NC limit, the expressions $a^2\bar{E}_W$ and $a^2\bar{I}(A:B)$ (given in eq(s).\eqref{EwUV},\eqref{HMiUV}) produces negative values for all possible values of $\frac{d}{a}$. The result therefore indicates that this cannot be a physical phase, which signals that the disconnected phase is the physical phase and therefore both $E_W$ and $I(A:B)$ are zero.\\
We now study the effect of UV/IR mixing on the EWCS and HMI. By UV/IR mixing we intend to investigate the effect of the UV cut-off on the IR result \cite{Karczmarek:2013xxa}. We compute the above quantities for two different values of the UV cut-off, namely, $au_b=10,20$.
\begin{figure}[!h]
	\centering
	\includegraphics[width=0.5\textwidth]{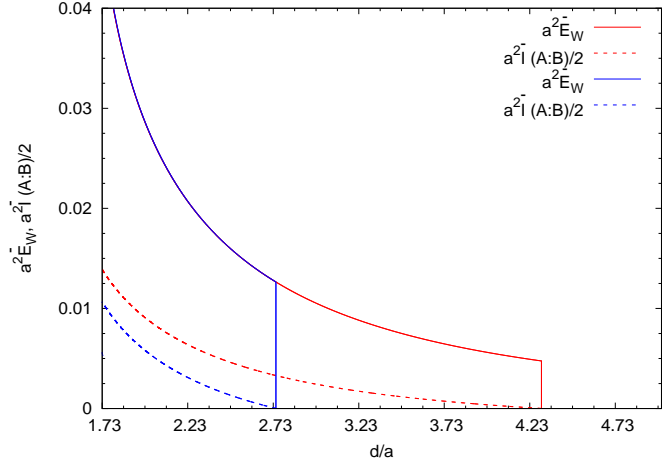}
	\caption{Effect of UV/IR mixing on EWCS and HMI . In the above plot, the red curves (both solid and dotted) correspond to $au_b=10$ and the blue curves (both solid and dotted) correspond to $au_b=20$. We have set $\frac{l}{a}=6$.}
	\label{figUVIR}
\end{figure}\\
From Fig.\eqref{figUVIR}, we can see clearly the prominent effect of the UV cut-off on the EWCS and HMI. The effect of the UV cut-off on these quantities is a signature of the UV/IR mixing \cite{Karczmarek:2013xxa}. In particular, we observe that for a fixed subsystem length $\frac{l}{a}$, the HMI and EWCS vanishes at a smaller value of separation $\frac{d}{a}$ for a larger cut-off value. This shows the sensitivity of the IR results on the UV cut-off. We make one more comment. In the IR domain, the HEE obeys an area law (eq.(\ref{Sdiv})) in contrast to the UV domain (eq.(\ref{ee2})) where it obeys a volume law. This is consistent with the fact that in the deep IR limit, the HEE reduces to the Bekenstein-Hawking entropy of the black hole. The fact that in the UV limit, the HEE obeys a volume law implies that it has an extensive behaviour unlike the IR behaviour.\\
Now we proceed to investigate the behaviour of mutual information in the deep UV region. Once again we consider two subsystems of equal length $l<l_{c}$ kept at a distance $d<l$, and also consider $2l+d < l_{c}$.
Substituting eq.(\ref{ee2}) in the
expression for mutual information eq.(\ref{mi}), we obtain 
\begin{eqnarray}\label{mi2}
a^{2}\bar{I}(A:B)&=&\left(-\left(\frac{d}{a}\right)(au_{b})^{3}+\left(\frac{3}{16}\right)(au_{b})\frac{6\left(\frac{l}{a}\right)^{3}+2\left(\frac{d}{a}\right)^{3}+12\left(\frac{l}{a}\right)^{2}\left(\frac{d}{a}\right)+6\left(\frac{l}{a}\right)\left(\frac{d}{a}\right)^{2}}{1+(\frac{1}{au_b})^{4}}\right.\nonumber\\
&+&\left.\left(\frac{9}{16 au_{b}}\right)\frac{30\left(\frac{l}{a}\right)^{5}+2\left(\frac{d}{a}\right)^{5}+80\left(\frac{l}{a}\right)^{4}\left(\frac{d}{a}\right)+40\left(\frac{l}{a}\right)^{2}\left(\frac{d}{a}\right)^{3}+10\left(\frac{l}{a}\right)\left(\frac{d}{a}\right)^{4}}{(1+(\frac{1}{au_b})^{4})^{2}}\right)~.\nonumber\\
\end{eqnarray}
From eq.(\ref{mi2}), one can see that the holographic mutual information is a divergent quantity in the deep UV regime. We know that the mutual information is in general UV insensitive. By this we mean that the divergent piece is independent of the length of the subsystem. But in this case, the divergent part in the HEE depends on the subsystem size. So in this case the divergent parts in the expression for mutual information do not cancel. Hence, the holographic mutual information is a divergent quantity in the deep UV regime and hence, there is no phase transition in the deep UV limit. This observation was made earlier in \cite{Karczmarek:2013xxa}. Further, we note that, in the deep UV limit, the evaluation of $a^2\bar{I}(A:B)$ (given in eq.\eqref{mi2}) produces negative values for all possible $\frac{d}{a}$. The result therefore indicates that this phase is not physical.
\section{Holographic entanglement entropy at finite temperature}\label{Sec4}
In this section, we discuss the information theoretic aspects of NC Yang-Mills theory at a finite temperature. The dual gravitational theory in this context is a black hole geometry associated with the following metric
\begin{eqnarray}\label{BH}
ds^{2}&=&R^{2}\bigg[-u^{2}f(u)dt^{2}+u^{2}dx_{1}^{2}+u^{2}h(u)(dx_{2}^{2}+dx_{3}^{2})+\frac{du^{2}}{u^{2}f(u)}\bigg]+R^{2}d\Omega_{5}^{2}~
\end{eqnarray}
where $f(u)=1-\left(\frac{u_H}{u}\right)^4$. In the lapse function $f(u)$, the position of the event horizon of the black hole is specified by $u_H$. Further it is related with the Hawking temperature $T_H$ of the black hole as $T_H = \frac{u_H}{\pi}$.\\
We now proceed to compute the HEE corresponding to a subsystem $A$ at the boundary by following the same set up we have used previously. This leads to the following integral for HEE$(S_{EE})$
\begin{equation}\label{tee}
S_{EE}=\frac{2R^{8}L^{2}Vol(\Omega_{5})}{4g_{s}^{2}G_{N}^{10}}\int_{-l/2}^{0}\frac{u^{2}}{\sqrt{h(u)}}\left(u^{2}h(u)+\frac{{u^{\prime}}^{2}}{u^{2}f(u)}\right)^{\frac{1}{2}}~dx_{2}~.
\end{equation}
The cyclicity of the coordinate $x_{2}$ gives the conserved Hamiltonian to be
\begin{equation}
\mathcal{H}=-\frac{u^{4}}{\left(u^{2}+\frac{{u^{\prime}}^{2}}{u^{2}f(u)h(u)}\right)^{\frac{1}{2}}}=constant(c^{\prime})=-u_{t}^{3}~.
\end{equation}
This now results in the following differential equation 
\begin{equation}\label{diff}
\frac{du}{dx_{2}}=\sqrt{u^{4}h(u)f(u)\left(\bigg(\frac{u}{u_{t}}\bigg)^{6}-1\right)}~.
\end{equation}
Now substituting eq.(\ref{diff}) in eq.(\ref{tee}) and using the boundary condition \ref{bc1} yields
\begin{eqnarray}\label{EE5}
a^2\bar{S}_{EE} =\frac{1}{2} (au_t)^2 \int_{\frac{au_t}{au_b}}^{1}\frac{\sqrt{p^{4}+(au_{t})^{4}}}{p^{5}\sqrt{1-p^{6}}\sqrt{1-\eta^4p^4}} dp~;~~\eta=\frac{au_H}{au_t}~~;~~p=\frac{au_{t}}{au}~.
\end{eqnarray}
We now make an approximation. We assume that the dimensionless parameter $\eta\ll1$ which can be identified as the low temperature approximation. By incorporating this, we have the following expression (for $au_{t}\le 1$)
\begin{eqnarray}\label{EE6}
a^2\bar{S}_{EE} &\simeq&\frac{1}{2} (au_t)^2\left[ \int_{\frac{au_t}{au_b}}^{1}\frac{\sqrt{p^{4}+(au_{t})^{4}}}{p^{5}\sqrt{1-p^{6}}} dp + \frac{\eta^4}{2}\int_{\frac{au_t}{au_b}}^{1}\frac{\sqrt{p^{4}+(au_{t})^{4}}}{p\sqrt{1-p^{6}}} dp\right]\nonumber\\
&\approx& a^{2}\bar{S}_{div}+\left(\sum_{n=1}^{\infty}\frac{1}{2\sqrt{\pi}}\frac{\Gamma(n+\frac{1}{2})}{\Gamma(n+1)} \left[\frac{1}{(6n-4)}+\frac{1}{(12n)}-\frac{1}{(6n-2)}\right]-\sum_{n=2}^{\infty}\frac{1}{4\sqrt{\pi}}\frac{\Gamma(n+\frac{1}{2})}{\Gamma(n+1)}\frac{1}{(6n-6)}\right)\nonumber\\
&\times&(au_t)^{(6n)}+\sum_{n=2}^{\infty}\frac{1}{4\sqrt{\pi}}\frac{\Gamma(n+\frac{1}{2})}{\Gamma(n+1)}\frac{(au_t)^{6}}{(6n-6)}+\sum_{n=0}^{\infty}\frac{1}{2\sqrt{\pi}}\frac{\Gamma(n+\frac{1}{2})}{\Gamma(n+1)}\frac{(au_{t})^{2}}{(6n-2)}\nonumber\\
&+&\left[\frac{1}{6}-\frac{(au_{t})^{6}}{24}-(au_{t})^{6}\frac{\Gamma(\frac{3}{2})\log(au_{t})}{4\sqrt{\pi}}\right]+\frac{\eta^{4}}{4}\bigg[(au_{t})^{2}\sum_{n=0}^{\infty}\frac{1}{\sqrt{\pi}}\frac{\Gamma(n+(1/2))}{\Gamma(n+1)}\frac{1}{6n+2}\nonumber\\
&+&\frac{(au_{t})^{4}}{2}\sum_{n=0}^{\infty}\frac{1}{\sqrt{\pi}}\frac{\Gamma(n+(1/2))}{\Gamma(n+1)}\frac{1}{6n-2}+\bigg(\sum_{n=0}^{\infty}\frac{1}{\sqrt{\pi}}\frac{\Gamma(n+(1/2))}{\Gamma(n+1)}\bigg[\frac{1}{2(6n+4)}-\frac{1}{6n+2}-\frac{1}{6n-2}\bigg]\nonumber\\
&+&\sum_{n=1}^{\infty}\frac{1}{\sqrt{\pi}}\frac{\Gamma(n+(1/2))}{\Gamma(n+1)}\frac{1}{6n}\bigg)(au_{t})^{6n+4}\bigg]~.	
\end{eqnarray}
It is a well known fact that computation of the above expression will lead to two terms, namely, a divergent term $S_{div}$ and a finite term $a^2\bar{S}_{EE}|^{finite}$ (as a function of the turning point $au_t$). By following same method we have introduced previously, one can compute the expression corresponding to $a^2\bar{S}_{EE}|^{finite}$.\\

\begin{figure}[!h]
	\begin{minipage}[t]{0.48\textwidth}
		\centering\includegraphics[width=\textwidth]{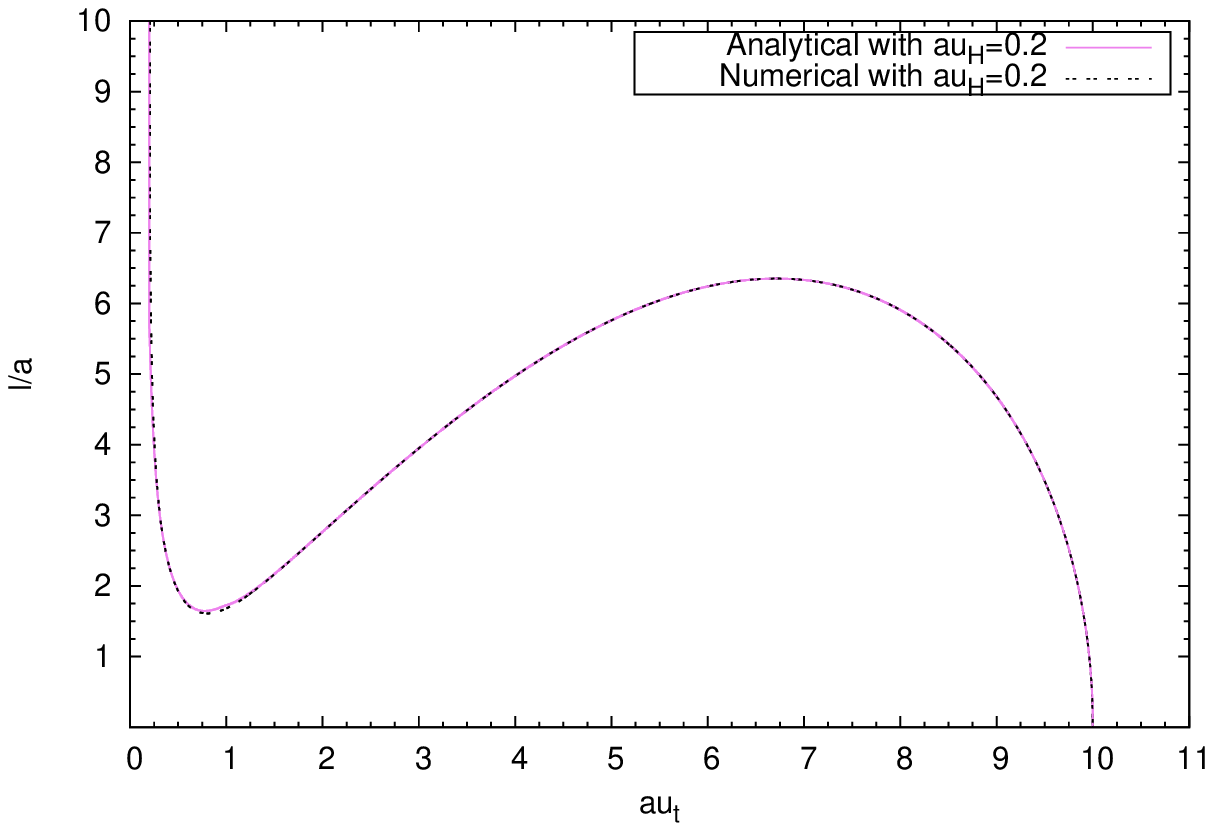}\\
		{\footnotesize Variation of $\frac{l}{a}$ with respect to $au_t$ with $au_H=0.2$.}
	\end{minipage}\hfill
	\begin{minipage}[t]{0.48\textwidth}
		\centering\includegraphics[width=\textwidth]{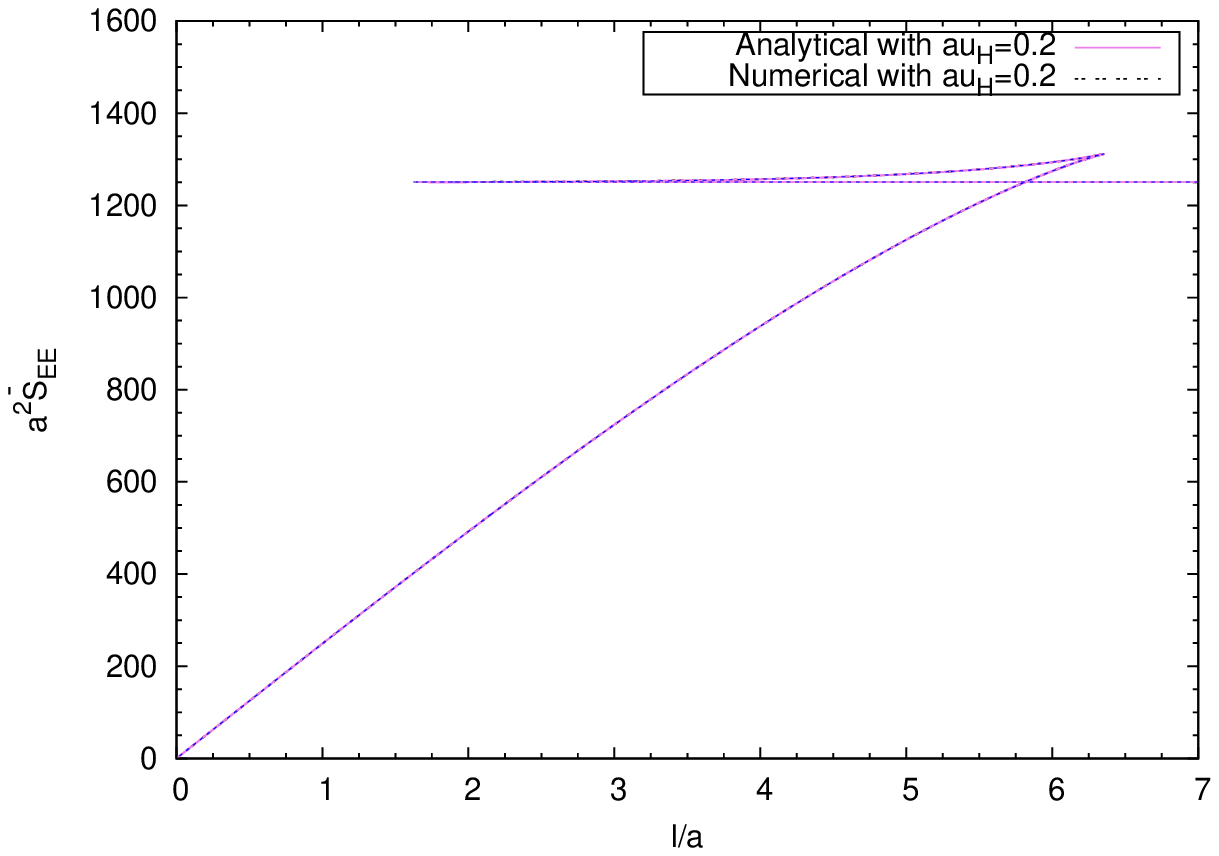}\\
		{\footnotesize Variation of the  $a^{2}\bar{S}_{EE}$ with respect to $\frac{l}{a}$ at $au_{H}=0.2$.}
	\end{minipage}
	\caption{In the left we have shown the variation of $\frac{l}{a}$ with respect to $au_t$, and in the right we have presented the variation of $a^{2}\bar{S}_{EE}$ with respect to $\frac{l}{a}$. In these plots we have set $au_{b}=10$.}
	\label{CFBH}
\end{figure}
\noindent The divergent piece in this scenario reads
\begin{eqnarray}\label{teediv}
a^{2}\bar{S}_{div}=\frac{1}{8}(au_{b})^{4}+\frac{\log(au_{b})}{4}+\frac{(a\pi T)^4}{4}\log(au_b)~.
\end{eqnarray} 
The above divergent piece is quite surprising as it contains a temperature dependent term $\frac{(a\pi T)^4}{4}\log(au_b)$. This can be interpreted as a unique feature of NC Yang-Mills theory as in general the divergent piece does not depend on the temperature. 
In the finite temperature context the critical value of the turning point depends on the choice of the temperature. On the other hand, in the small $au_t$ approximation, $a^2\bar{S}_{EE}|^{finite}$ reads
\begin{eqnarray}\label{Se}
a^2\bar{S}_{EE}|^{finite} &\simeq&	\left[\frac{1}{6}-\frac{\sqrt{\pi}}{4}\frac{\Gamma(2/3)}{\Gamma(1/6)}(au_t)^2+\frac{1}{48}\left(3+\log 4-6\log(au_t)\right)(au_t)^6\right]\nonumber\\
&-&\eta^4\left[\frac{(au_t)^4}{32}+\frac{\sqrt{\pi}(au_t)^6}{16}\frac{\Gamma(2/3)}{\Gamma(1/6)}-\frac{\sqrt{\pi}(au_t)^2}{8}\frac{\Gamma(4/3)}{\Gamma(5/6)}+\frac{(au_t)^{10}}{240}\right]~.
\end{eqnarray}
It is to be noted that the above expression has been written under the condition $au_H\ll au_t\le 1$. This approximation leads to the leading order NC correction at low temperature. For $au_{t}\ge 1$, we have 
\begin{eqnarray}
a^{2}\bar{S}_{EE}&=&\frac{1}{8}\left((au_{b})^{4}-(au_{t})^{4}\right)-\frac{1}{4}\ln\left(\frac{au_{t}}{au_{b}}\right)+\frac{1}{4}\sum_{m=2}^{\infty}\frac{\sqrt{\pi}}{\Gamma(m+1)\Gamma(\frac{3}{2}-m)}\frac{1}{(au_{t})^{4(m-1)}}\frac{\left(1-\left(\frac{au_{t}}{au_{b}}\right)^{4(m-1)}\right)}{4(m-1)}\nonumber\\
&+&\frac{1}{4}\sum_{n=1}^{\infty}\sum_{m=0}^{\infty}\frac{\Gamma(n+\frac{1}{2})}{\Gamma(m+1)\Gamma(\frac{3}{2}-m)\Gamma(n+1)}\frac{1}{\sqrt{\pi}}\frac{1}{(au_{t})^{4(m-1)}}\frac{\left(1-\left(\frac{au_{t}}{au_{b}}\right)^{6n+4(m-1)}\right)}{6n+4(m-1)}\nonumber\\
&+&\frac{(au_{H})^{4}}{4}\left[\ln(au_{b})+\sum_{m=1}^{\infty}\frac{\Gamma(3/2)}{\Gamma(m+1)\Gamma(\frac{3}{2}-m)}\frac{1}{(au_{t})^{4m}}\frac{\left(1-\left(\frac{au_{t}}{au_{b}}\right)^{4m}\right)}{4m}\right.\nonumber\\
&+&\left.\sum_{n=1}^{\infty}\sum_{m=0}^{\infty}\frac{\Gamma(\frac{3}{2})}{\Gamma(m+1)\Gamma(\frac{3}{2}-m)}\frac{\Gamma(n+1)}{\Gamma(n+\frac{1}{2})\sqrt{\pi}}\frac{1}{(au_{t})^{4m}}\frac{\left(1-\left(\frac{au_{t}}{au_{b}}\right)^{6n+4m}\right)}{6n+4m}\right]~.
\end{eqnarray}
Note that one can recover the expression for HEE for $au_{t}\ge 1$ at $T=0$ given in eq.(\ref{see1}) by setting $u_{H}=0$ in the above expression.
We now write down this expression as a function of the subsystem size $\frac{l}{a}$. To do this we note that the subsystem size in terms of the bulk coordinate can be written as 
\begin{equation}\label{lbh}
\frac{l}{a}=\frac{2}{au_{t}}\int_{\frac{au_{t}}{au_{b}}}^{1}\frac{p\sqrt{p^{4}+(au_{t})^{4}}}{\sqrt{1-p^{6}}\sqrt{1-\eta^{4}p^{4}}}~dp~.
\end{equation}
The integral for subsystem size $\frac{l}{a}$ in the low temperature approximation reads (for $au_{t}\le 1$)
\begin{eqnarray}\label{lbh5}
\frac{l}{a}&\simeq&\frac{2}{(au_{t})}\left[\int_{\frac{au_{t}}{au_{b}}}^{1} dp \frac{p\sqrt{p^{4}+(au_{t})^{4}}}{\sqrt{1-p^{6}}}+\frac{\eta^4}{2}\int_{\frac{au_{t}}{au_{b}}}^{1} dp \frac{p^5\sqrt{p^{4}+(au_{t})^{4}}}{\sqrt{1-p^{6}}}\right]\nonumber\\
&\approx&\frac{\sqrt{\pi}}{2(au_t)}\frac{\Gamma(\frac{5}{3})}{\Gamma(\frac{7}{6})}-(au_{t})^{3}\ln(au_{t})+(au_t)^3\sum_{n=1}^{\infty}\frac{1}{\sqrt{\pi}}\frac{\Gamma(n+\frac{1}{2})}{\Gamma(n+1)}\frac{1}{(6n)}-\sum_{n=1}^{\infty}\frac{1}{\sqrt{\pi}}\frac{\Gamma(n+\frac{1}{2})}{\Gamma(n+1)}\frac{(au_t)^{(6n+3)}}{(6n)}\nonumber\\
&+&\left(\sum_{n=0}^{\infty}\frac{2}{\sqrt{\pi}}\frac{\Gamma(n+\frac{1}{2})}{\Gamma(n+1)}\left[\frac{1-(1/au_{b})^{6n+2}}{(6n+2)}-\frac{1}{(6n+4)}+\frac{1-(1/au_{b})^{6n+6}}{2(6n+6)}\right]\right)(au_t)^{(6n+3)}\nonumber\\
&+&\eta^{4}\bigg[\frac{\sqrt{\pi}}{8}\frac{\Gamma(7/3)}{\Gamma(11/6)}\frac{1}{au_{t}}+\frac{\sqrt{\pi}}{8}\frac{\Gamma(5/3)}{\Gamma(7/6)}(au_{t})^{3}+\sum_{n=0}^{\infty}\frac{1}{\sqrt{\pi}}\frac{\Gamma(n+(1/2))}{\Gamma(n+1)}\bigg(\frac{1-(1/au_{b})^{6n+6}}{6n+6}\nonumber\\
&-&\frac{1}{6n+8}-\frac{1}{2(6n+4)}+\frac{1}{2}\frac{1-(1/au_{b})^{6n+10}}{6n+10}\bigg)(au_{t})^{6n+7}\bigg]~.
\end{eqnarray}
Once again we compare our analytical result (given below) with the numerical one. In the left panel of fig.(\ref{CFBH}), we have graphically represented this comparison for  $au_{H}=0.2$. The analytical expression of $\left(\frac{l}{a}\right)$ has been obtained from eq.(\ref{lbh5}) by following the technique introduced for the zero temperature case. From this figure, it can be seen that the numerical and analytical results are in good agreement with each other. We now proceed to compute the value of $(au_{t})_{c}$ at which the length scale $\left(\frac{l}{a}\right)_{min}$ appears.
In the deep IR limit, that is, $au_{t}\ll 1$, eq.(\ref{lbh}) leads to the following expression
\begin{equation}\label{lbhir}
\left(\frac{l}{a}\right)_{deep~ IR}\approx\frac{2\sqrt{\pi}}{(au_{t})}\frac{\Gamma(2/3)}{\Gamma(1/6)}+\frac{(au_{H})^{4}}{(au_{t})^{5}}\frac{\sqrt{\pi}}{8}\frac{\Gamma(7/3)}{\Gamma(11/6)}~.
\end{equation}
In the deep noncommutative limit, that is, $(au_{t})\gg 1$, and $au_{t}\ll au_{b}$ \cite{Barbon:2008ut}, one can obtain the following expression from eq.(\ref{lbh})
\begin{equation}\label{lbhuv}
\left(\frac{l}{a}\right)_{deep~NC}\approx\frac{\sqrt{\pi}}{3}\frac{\Gamma(1/3)}{\Gamma(5/6)}(au_{t})+\frac{1}{3}\frac{(au_{H})^{4}}{(au_{t})^{3}}~.
\end{equation}
In the computation of the above expression, we have considered terms upto $\mathcal{O}((au_{H})^{4})$. By equating the results given in eq.(\ref{lbhir})and eq.(\ref{lbhuv}), we obtain an equation of the form
\begin{eqnarray}
2\sqrt{\pi}\frac{\Gamma(2/3)}{\Gamma(1/6)}(au_{t})^{4}-\frac{\sqrt{\pi}}{3}\frac{\Gamma(1/3)}{\Gamma(5/6)}(au_{t})^{6}-(au_{H})^{4}\left(\frac{(au_{t})^{2}}{3}-\frac{\sqrt{\pi}}{8}\frac{\Gamma(7/3)}{\Gamma(11/6)}\right)=0~.
\end{eqnarray}
Now if we choose $au_{H}=0.2$, the value of $(au_{t})_{c}$ is obtained to be $(au_{t})_{c}\approx0.784$. Furthermore, the value of $\left(\frac{l}{a}\right)_{min}$ is found to be  $\left(\frac{l}{a}\right)_{min}\approx1.649$. We conclude that in the finite temperature scenario, the values of $(au_{t})_{c}$ and  $\left(\frac{l}{a}\right)_{min}$ depend on the choice of $(au_{H})$. 
We can now follow the same procedure we have performed previously and compute the above integrals. The above integrals can be easily computed by the technique we have introduced in this paper. In the limit $au_H\ll au_t\ll1$, the expression reads
\begin{eqnarray}\label{lbhapp}
\frac{l}{a}&\simeq& \frac{\sqrt{\pi}}{2(au_{t})}\frac{\Gamma(5/3)}{\Gamma(7/6)}\left[1+\frac{1}{3\sqrt{\pi}}\frac{\Gamma(7/6)}{\Gamma(5/3)}\left(4+\log4-6\log(au_t)\right)(au_t)^4\right]\nonumber\\
&+&\eta^4\left[\frac{\sqrt{\pi}(au_t)^3}{8}\frac{\Gamma(5/3)}{\Gamma(7/6)}+\frac{\sqrt{\pi}}{8(au_t)}\frac{\Gamma(7/3)}{\Gamma(11/6)}-\frac{(au_t)^7}{30}\right]~.
\end{eqnarray}
By following the perturbative approach, we can express the turning point $au_t$ in terms of the subsystem size $\frac{l}{a}$, for large $\frac{l}{a}$. This reads
\begin{eqnarray}\label{tbh}
au_t &=& \left[\frac{\alpha_0}{(\frac{l}{a})}+\frac{\alpha_1}{(\frac{l}{a})^5}+\frac{\alpha_0^4\log(\frac{l}{a})}{(\frac{l}{a})^5}\right] + \frac{(a\pi T)^4}{4}\left[\frac{\alpha_0}{(\frac{l}{a})}+\frac{1}{\alpha_0^4}\frac{\Gamma(7/3)}{\Gamma(11/6)}\left(\frac{l}{a}\right)^3-\frac{4}{30}\frac{\alpha_0^4}{(\frac{l}{a})^5}\right]\nonumber\\
&\equiv& l_1 + \frac{(a\pi T)^4}{4} l_2
\end{eqnarray}
where $l_1$ and $l_2$ are given by
\begin{eqnarray}
l_1 = \frac{\alpha_0}{(\frac{l}{a})}+\frac{\alpha_1}{(\frac{l}{a})^5}+\frac{\alpha_0^4\log(\frac{l}{a})}{(\frac{l}{a})^5};~~l_2 = \frac{\alpha_0}{(\frac{l}{a})}+\frac{1}{\alpha_0^4}\frac{\Gamma(7/3)}{\Gamma(11/6)}\left(\frac{l}{a}\right)^3-\frac{4}{30}\frac{\alpha_0^5}{(\frac{l}{a})^5}\nonumber~.
\end{eqnarray}
For $au_{t}\ge 1$, we have
\begin{eqnarray}\label{Lt}
\frac{l}{a}&=&\sum_{m=0}^{\infty}\sum_{n=0}^{\infty}\frac{\Gamma(n+\frac{1}{2})}{\Gamma(m+1)\Gamma(\frac{3}{2}-m)\Gamma(n+1)}\frac{1}{(au_{t})^{4m-1}}\frac{\left(1-\left(\frac{au_{t}}{au_{b}}\right)^{6n+4m+2}\right)}{6n+4m+2}\nonumber\\
&+&\eta^{4}\sum_{m=0}^{\infty}\sum_{n=0}^{\infty}\frac{\Gamma(n+\frac{1}{2})}{\Gamma(n+1)\Gamma(m+1)\Gamma(\frac{3}{2}-m)}\frac{1}{(au_{t})^{4m-1}}\frac{\left(1-\left(\frac{au_{t}}{au_{b}}\right)^{6n+4m+6}\right)}{6n+4m+6}~.
\end{eqnarray}
From the above expression it can be observed that a temperature dependent term has appeared due to the thermal excitation we have considered. It is to be noted that in the above computation we have kept terms upto $\mathcal{O}\left(T^4\right)$. We now substitute the above expression for the turning point $au_t$ (eq.(\ref{tbh})) in eq.\eqref{Se} and obtain the following expression (for $au_{t}<<1$, large $\frac{l}{a}$)
\begin{eqnarray}\label{EET}
a^2\bar{S}_{EE}|^{finite}\left(\frac{l}{a}\right) = a^2\bar{S}_{EE}\left(\frac{l}{a}\right)|^{finite} + \frac{\left(a\pi T\right)^4}{4}\Delta \left(a^2\bar{S}_{EE}\right)~.
\end{eqnarray}
In the above expression, $a^2\bar{S}_{EE}\left(\frac{l}{a}\right)|^{finite}$ represents the finite piece of HEE corresponding to $T=0$ case and the expression is given in eq.\eqref{EEl}. Further, $\Delta \left(a^2\bar{S}_{EE}\right)$ represents the change in HEE due to the thermal excitation, with the following expression
\begin{eqnarray}
\Delta \left(a^2\bar{S}_{EE}\right) &=& \frac{\sqrt{\pi}}{2l_1^2}\frac{\Gamma(4/3)}{\Gamma(5/6)}-\frac{\sqrt{\pi}}{2}\frac{\Gamma(2/3)}{\Gamma(1/6)}l_1l_2+\frac{(3+\log 4)}{8}l_1^5l_2-\frac{\sqrt{\pi}}{4}\frac{\Gamma(2/3)}{\Gamma(1/6)}l_1^2-\frac{l_1^6}{60}\nonumber\\
&-&\frac{1}{8}l_1^5l_2\log(l_1)-\frac{l_1^5l_2}{8}-\frac{1}{8}~.	
\end{eqnarray} 
It is to be noted that in the above expression we have kept terms upto $\mathcal{O}\left(T^4\right)$. The change in the HEE due to thermal excitation plays a crucial role in context of entanglement thermodynamics\cite{Bhattacharya:2012mi,Karar:2018ecr,Saha:2019ado,Saha:2020fon,Karar:2020cvz}. In the right panel of fig.\eqref{CFBH}, we have graphically represented  $a^2\bar{S}_{EE}|^{finite}\left(\frac{l}{a}\right)$ as a function of $(l/a)$ (given in eq.\eqref{EET}). The presence of the length scale $(\frac{l}{a})_{min}$ can be noticed from the plot in the left panel. One can also observe the effect of temperature from the plot in the right panel.

\noindent In order to probe the effect of noncommutativity we now compute the HEE at the deep IR limit, which corresponds to the commutative Yang-Mills theory. This reads
\begin{eqnarray}
\left(a^2\bar{S}_{EE}|^{finite}\left(\frac{l}{a}\right)\right)_{deep~ IR} = \left(a^2\bar{S}_{EE}|^{finite}\left(\frac{l}{a}\right)\right)_{deep~ IR}^{T=0} + \frac{\left(a\pi T\right)^4}{4}\Delta \left(a^2\bar{S}_{EE}\right)_{deep~IR}\nonumber\\	
\end{eqnarray}
where $\left(a^2\bar{S}_{EE}|^{finite}\left(\frac{l}{a}\right)\right)_{deep~ IR}^{T=0}$ represents the temperature independent piece (given in eq.\eqref{EE4}) and\\
$\Delta \left(a^2\bar{S}_{EE}\right)_{deep~IR}$ probes the change in HEE (due to thermal excitation) with the following expression
\begin{eqnarray}
\Delta \left(a^2\bar{S}_{EE}\right)_{deep~IR} =\frac{1}{\sqrt{\pi}} \left(\frac{\Gamma(7/6)}{\Gamma(5/3)}\right)^2\left(\frac{\Gamma(1/3)}{\Gamma(5/6)}\right)\left(\frac{l}{a}\right)^2\left[\frac{4}{3}-\frac{432}{5^5}\left(\frac{\Gamma(7/6)}{\Gamma(5/3)}\right)\right]~.
\end{eqnarray} 
All the analysis done above is valid for a strip length $l$ larger than some critical length $l_{c}$ which we shall see now. It has been observed in \cite{Karczmarek:2013xxa} that to study the surfaces anchored on small strips, $u(x_{2})$ has to be expanded in a power series of $x_{2}$ (see eq.(\ref{a})).
Substituting this power series in eq.(\ref{diff}) and using $u(x_{2}=0)=u_{t}$ and the boundary condition given in eq.(\ref{bc1}), we get the profile of the RT surface to be
\begin{equation}
u(x_{2})=u_{t}+\frac{3}{2}\frac{u_{t}^{3}}{1+(au_{t})^{4}}\left(1-\bigg(\frac{u_{H}}{u_{t}}\bigg)^{4}\right)x_{2}^{2}+... ~.
\end{equation}
Now putting $x_{2}=\frac{l}{2}$, we get 
\begin{equation}
u_{b}=u_{t}+\frac{3}{8}\frac{u_{t}^{3}}{[1+(au_{t})^{4}]}\left(1-\bigg(\frac{u_{H}}{u_{t}}\bigg)^{4}\right)l^{2}+\mathcal{O}\left((l/a)^{4}\right)~.
\end{equation}
Now substituting the above result in eq.(\ref{tee}), we get the entanglement entropy as 
\begin{eqnarray}\label{tee2}
a^{2}\bar{S}_{EE}&=&\frac{a^{2}}{2}\left(\frac{l}{\epsilon^{3}}-\frac{3}{8}\frac{l^{3}}{\epsilon a^{4}}\frac{(1-(\epsilon u_{H})^{4})}{[1+(\epsilon/a)^{4}]}-\frac{9}{8}\frac{l^{5}\epsilon}{a^{8}}\frac{[1+3(\epsilon u_{H})^{4}/10]}{[1+(\epsilon/a)^{4}]^{2}}
+\mathcal{O}\left((l/a)^{7},\epsilon^{6}\right)\right)\nonumber\\
&=&\frac{1}{2}\left((au_{b})^{3}\frac{l}{a}-\frac{3}{8}\left(\frac{l}{a}\right)^{3}\frac{(au_{b})(1-(au_{H}/au_{b})^{4})}{\left(1+(1/au_{b})^{4}\right)}-\frac{9}{8}\left(\frac{l}{a}\right)^{5}\frac{(1+\frac{3}{10}(au_{H}/au_{b})^{4})}{au_{b}\left(1+(1/au_{b})^{4}\right)^{2}}\right)\nonumber\\
&+&\mathcal{O}\left((l/a)^{7},\epsilon^{6}\right)~.
\end{eqnarray}
This result matches with the zero temperature result eq.{\ref{ee2}} when $u_{H}=0$. We observe that temperature arises only in the finite terms of the above expression. Now to obtain the critical length $l_{c}$, we equate the leading order divergent terms appearing in eq.(\ref{tee2}) and eq.(\ref{teediv}). This results in \cite{Fischler:2013gsa}
\begin{equation}\label{lcth}
l_{c}=\frac{a^{2}u_{b}}{2}+\left(\frac{1+(a\pi T)^{4}}{a^{2}u_{b}^{3}}\right)\ln(au_{b})~.
\end{equation}\section{EWCS at finite temperature}\label{Sec5}
In this section we proceed to compute the EWCS in the finite temperature scenario. Similar to the zero temperature case, we again consider two disjoint subsystems, namely, $A$ and $B$ (along the direction $x_2$) with length $l$ and separated by a distance $d$. The induced metric on constant $x_2$ hypersurface at finite temperature reads
\begin{eqnarray}\label{IndMT}
ds_{ind}^2&=&R^{2}\bigg[u^{2}dx_{1}^{2}+u^{2}h(u)dx_{3}^{2}+\frac{du^{2}}{u^{2}f(u)}\bigg]+R^{2}d\Omega_{5}^{2}~.
\end{eqnarray}
The above mentioned induced metric leads to the following expression for EWCS
\begin{eqnarray}\label{EWT}
a^{2}\bar{E}_W &=&\frac{a^{2}}{4}\int_{u_t(2l+d)}^{u_t(d)}\frac{u\sqrt{1+a^4u^4}}{\sqrt{f(u)}} ~du~\nonumber\\
&=&\frac{a^{2}}{16}\bigg[(au_{t}(d))^{2}\sqrt{1+(au_{t}(d))^{4}}\sqrt{1-\left(\frac{au_{H}}{au_{t}(d)}\right)^{4}}
+(1+(au_{H})^{4})\sinh[-1](\frac{(au_{t}(d))^{2}\sqrt{1-\left(\frac{au_{H}}{au_{t}(d)}\right)^{4}}}{\sqrt{1+(au_{H})^{4}}})\nonumber\\
&-&(au_{t}(2l+d))^{2}\sqrt{1+(au_{t}(2l+d))^{4}}\sqrt{1-\left(\frac{au_{H}}{au_{t}(2l+d)}\right)^{4}}\nonumber\\
&-&(1+(au_{H})^{4})\sinh[-1](\frac{(au_{t}(2l+d))^{2}\sqrt{1-\left(\frac{au_{H}}{au_{t}(2l+d)}\right)^{4}}}{\sqrt{1+(au_{H})^{4}}})\bigg]~~.
\end{eqnarray}
For $au_{t}\le 1$, the above expression simplifies to 
\begin{eqnarray}
a^{2}\bar{E}_{W}&=&\frac{1}{8}\left[(au_{t}(d))^{2}-(au_{t}(2l+d))^{2}\right]+\frac{1}{32}\left[(au_{t}(d))^{6}-(au_{t}(2l+d)^{6})\right]\nonumber\\
&+&\frac{(au_{H})^{4}}{16}\left[\frac{1}{(au_{t}(2l+d))^{2}}-\frac{1}{(au_{t}(d))^{2}}\right]+\frac{3}{64}(au_{H})^{4}\left[(au_{t}(2l+d))^{2}-(au_{t}(d))^{2}\right]~~.
\end{eqnarray}
In the deep IR (commutative) limit, the expression of EWCS in terms of the subsystem size is given by
\begin{eqnarray}\label{ewcstir}
a^{2}\bar{E}_{W}|_{deep~ IR}&=&\frac{1}{8}\left[\left(\frac{\alpha_{0}}{\left(\frac{d}{a}\right)}+\left(\frac{d}{a}\right)^{2}A~(a\pi T)^{4}\right)^{2}-\left(\frac{\alpha_{0}}{\left(\frac{2l+d}{a}\right)}+\left(\frac{2l+d}{a}\right)^{2}A~(a\pi T)^{4}\right)^{2}\right]\nonumber\\
&+&\frac{(a\pi T)^{4}}{16}\left[\frac{1}{\left(\frac{\alpha_{0}}{\left(\frac{2l+d}{a}\right)}+\left(\frac{2l+d}{a}\right)A~(a\pi T)^{4}\right)^{2}}-\frac{1}{\left(\frac{\alpha_{0}}{\left(\frac{d}{a}\right)}+\left(\frac{d}{a}\right)A~(a\pi T)^{4}\right)^{2}}\right]
\end{eqnarray}
where $A$ is given by
\begin{equation*}
A=\frac{\Gamma(\frac{7}{3})}{12\sqrt{\pi}\Gamma(\frac{11}{6})}\left(\frac{\Gamma(\frac{1}{6})}{\Gamma(\frac{2}{3})}\right)^{2}~.
\end{equation*}
Further, the holographic mutual information at finite temperature reads (for $au_{t} \le 1$)
\begin{eqnarray}
a^2\bar{I}(A:B) &=& 2a^2\bar{S}_{EE}\left(\frac{l}{a}\right)-a^2\bar{S}_{EE}\left(\frac{d}{a}\right)-a^2\bar{S}_{EE}\left(\frac{2l+d}{a}\right)
\end{eqnarray}
where $a^{2}\bar{S}_{EE}(l/a)$ is given by eq.(\ref{EE6}). Once again to compute $a^{2}\bar{I}(A:B)$, we need to use the analytical expressions in eq.(s)(\ref{EE6},\ref{lbh5}) .\\ 
We now incorporate the expression given in eq.\eqref{EE6}, in order to understand the effect of temperature on $a^2\bar{I}(A:B)$. This leads to the following
\begin{eqnarray}
a^2\bar{I}(A:B) &=& a^2\bar{I}(A:B)|^{T=0}+	\frac{(a\pi T)^4}{4}\Delta\left(a^2\bar{I}(A:B)\right)~.
\end{eqnarray}
In the above expression, $a^2\bar{I}(A:B)|^{T=0}$ represents the temperature independent piece (given in eq.\eqref{HMI}) and $\Delta\left(a^2\bar{I}(A:B)\right)$ represents the change in HMI due to thermal excitation. This temperature dependent piece has the following expression
\begin{eqnarray}
\Delta\left(a^2\bar{I}(A:B)\right) = \left[2\Delta \left(a^2\bar{S}_{EE}\left(\frac{l}{a}\right)\right)-\Delta \left(a^2\bar{S}_{EE}\left(\frac{d}{a}\right)\right)-\Delta \left(a^2\bar{S}_{EE}\left(\frac{2l+d}{a}\right)\right)\right]~.
\end{eqnarray}
On the other hand, in the deep IR limit (commutative limit), the EWCS is obtained to be
\begin{eqnarray}
a^2\bar{E}_W|_{deep~IR} = a^2\bar{E}_W|_{deep~IR}^{T=0} + \frac{(a\pi T)^4}{4} \Delta\left(a^2\bar{E}_W\right)_{deep~IR}
\end{eqnarray} 
where the temperature independent piece $a^2\bar{E}_W|_{deep IR}^{T=0}$ is given in eq.\eqref{EwIR} and  $\Delta\left(a^2\bar{E}_W\right)_{deep~IR}$ represents the change in EWCS due to thermal excitation in the deep IR limit. Using eq.(\ref{ewcstir}), $\Delta\left(a^2\bar{E}_W\right)_{deep~IR}$ is given by
\begin{eqnarray}
\Delta\left(a^2\bar{E}_W\right)_{deep~IR}&=&A~\alpha_{0}\left(\left(\frac{d}{a}\right)-\left(\frac{2l+d}{a}\right)\right)
+\frac{1}{4}\left(\frac{1}{\left(\frac{\alpha_{0}}{\left(\frac{2l+d}{a}\right)}\right)^{2}}-\frac{1}{\left(\frac{\alpha_{0}}{\left(\frac{d}{a}\right)}\right)^{2}}\right)~.
\end{eqnarray}
We now graphically represent our computed results in order to have a better understanding of the above discussion.
\begin{figure}[!h]
	\begin{minipage}[t]{0.48\textwidth}
		\centering\includegraphics[width=\textwidth]{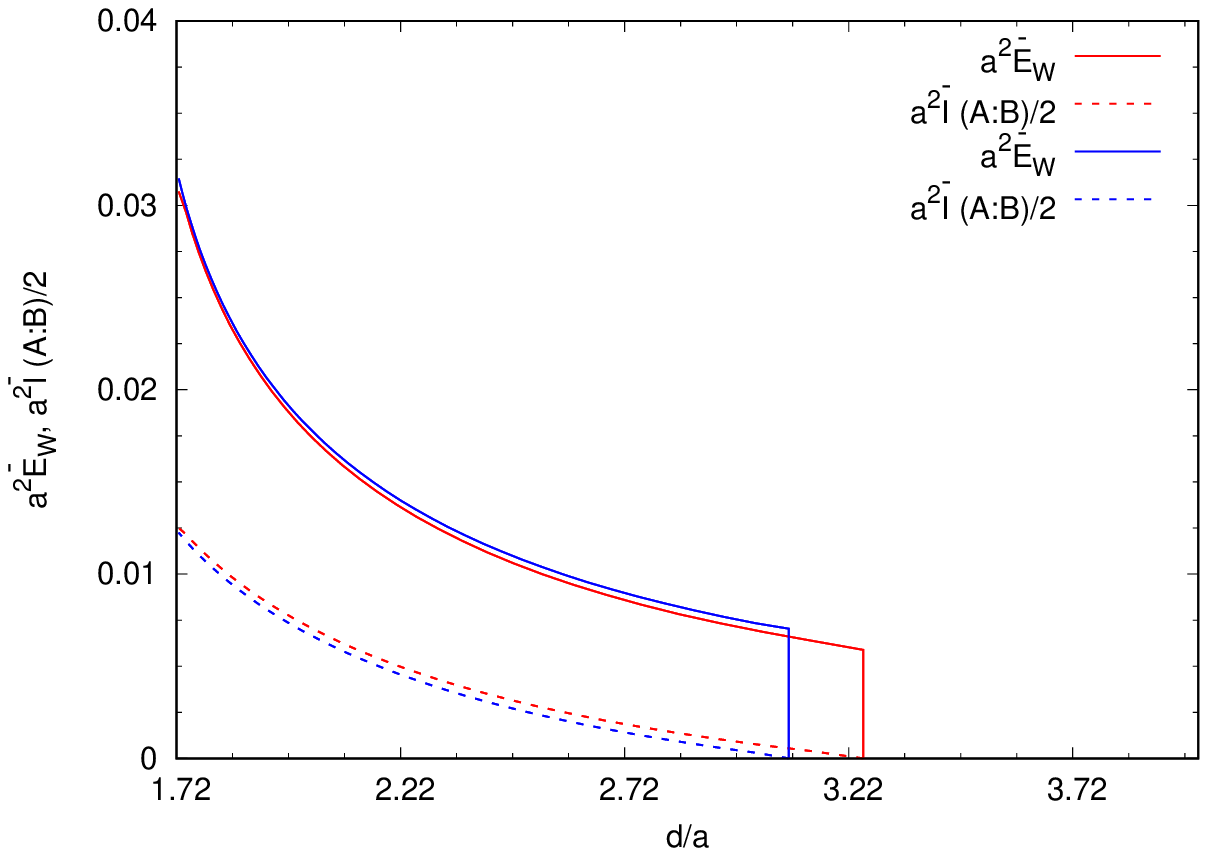}\\
		{\footnotesize Noncommutative Yang-Mills theory}
	\end{minipage}\hfill
	\begin{minipage}[t]{0.48\textwidth}
		\centering\includegraphics[width=\textwidth]{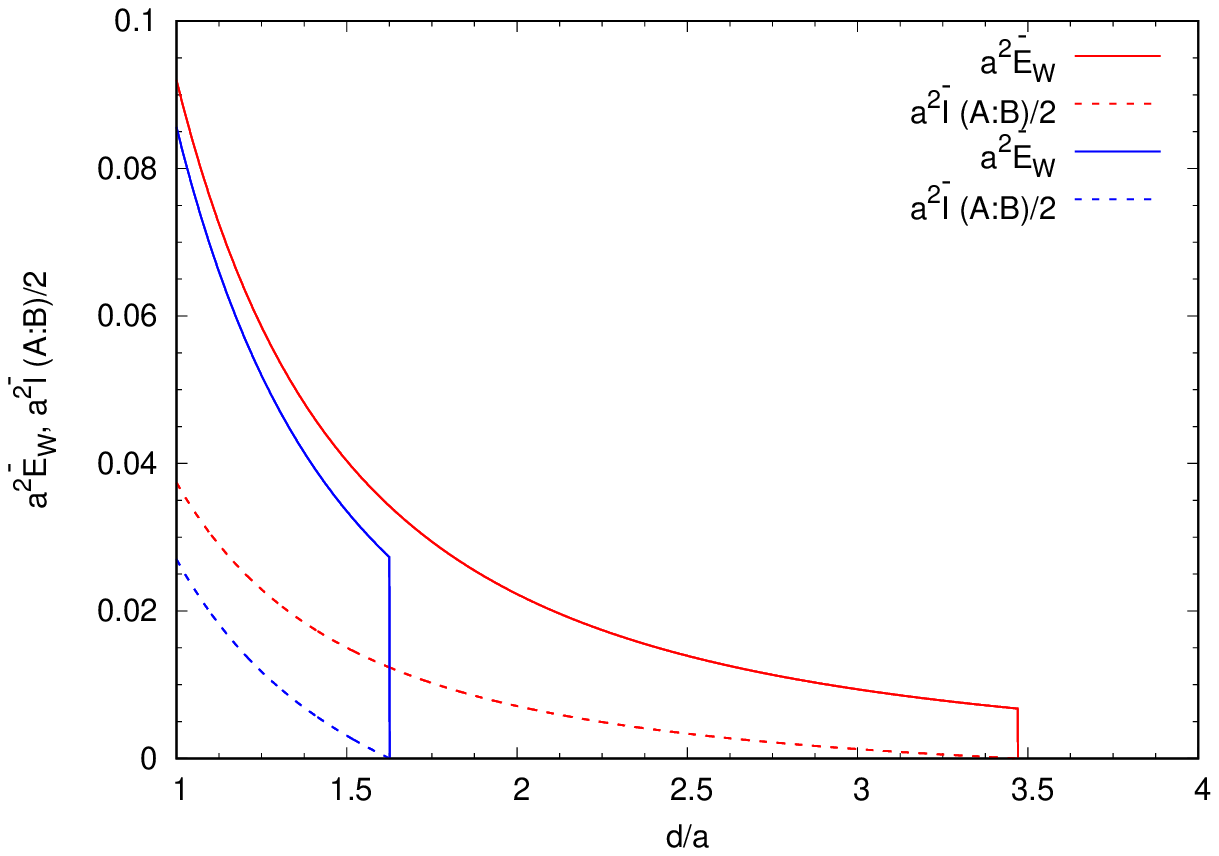}\\
		{\footnotesize Commutative (deep IR) Yang-Mills theory}
	\end{minipage}
	\caption{Effect of noncommutativity on EWCS and HMI at finite temperature with $aT=\frac{0.1}{\pi}$ (red) and $aT=\frac{0.2}{\pi}$ (blue) (we have set $\frac{l}{a}=6$) is shown in the left panel. The curve depict the analytical results. The value of $au_{b}=10$ for which eq.(\ref{lcth}) gives $\frac{l_{c}}{a}=5.0023$ for $aT=\frac{0.1}{\pi}$, and $\frac{l_{c}}{a}=5.00232$ for $aT=\frac{0.2}{\pi}$. The right panel depict the corresponding results in the deep IR limit.}
	\label{fig3}
\end{figure}\\
In Fig.\eqref{fig3}, we have graphically represented the effect of noncommutativity on EWCS and HMI at finite temperature. We have chosen two different values of temperature, namely, $aT=\frac{0.1}{\pi}$ (red) and $aT=\frac{0.2}{\pi}$ (blue). We observe that in presence of noncommutativity, the connected to disconnected phase transition of the RT surface $\Gamma_{AB}^{min}$ occurs at a higher value of the critical separation length $\left(\frac{d}{a}\right)_c$. Further it can be observed that the value of the critical separation length $\left(\frac{d}{a}\right)_c$ decreases with the increase in the temperature. The above plots also suggest that $a^2E_W\geq \frac{1}{2}a^2I(A:B)$ for all valid temperatures. Once again we choose the lower limit of the $d/a$ axis (for the noncommutative case in the left panel of Fig.\eqref{fig3}) to be $d/a=1.649$ (for $T=\frac{0.2}{\pi}$) so that $au_t(d/a)\leq1$. \\
\begin{figure}[!h]
	\centering
	\includegraphics[width=0.5\textwidth]{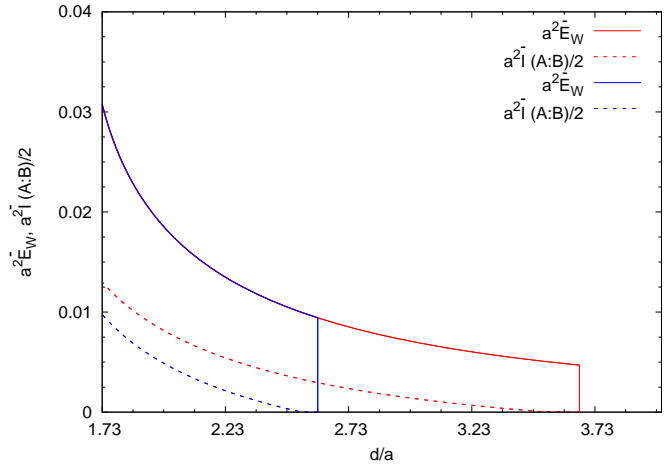}
	\caption{Effect of UV/IR mixing on EWCS and HMI at finite temperature $aT=\frac{0.1}{\pi}$. In the above plot, the red curves (both solid and dotted) correspond to $au_b=10$ and the blue curves (both solid and dotted) correspond to $au_b=20$. We have set $\frac{l}{a}=6$.}
	\label{figTUVIR}
\end{figure}\\
In Fig.\eqref{figTUVIR}, we once again probe the effect of UV/IR mixing on the EWCS and HMI, at a finite temperature. Similar to the zero temperature scenario, we observe that for a fixed subsystem length $\frac{l}{a}$, the HMI and EWCS vanishes at a smaller value of separation $\frac{d}{a}$ for a larger cut-off value. This again shows the sensitivity of the IR results on the UV cut-off.\\
Now we will consider again the behaviour of the holographic mutual information (HMI) below the critical length $l_{c}$ given in eq.(\ref{lcth}). Following the same procedure to obtain the holographic mutual information below the critical length for zero temperature, we obtain the holographic mutual information at finite temperature using eq.(\ref{tee2}). In this case it reads
\begin{eqnarray}\label{hmit}
a^{2}\bar{I}(A:B)&=&\frac{1}{2}\bigg[-2\left(\frac{d}{a}\right)(au_{b})^{3}+\frac{3}{8}(au_{b})\frac{\left(1-\left(\frac{au_{H}}{au_{b}}\right)^{4}\right)}{\left(1+(\frac{1}{au_b})^{4}\right)}\bigg\{6\left(\frac{l}{a}\right)^{3}+2\left(\frac{d}{a}\right)^{3}\nonumber\\
&+&12\left(\frac{l}{a}\right)^{2}\left(\frac{d}{a}\right)+6\left(\frac{l}{a}\right)\left(\frac{d}{a}\right)^{2}\bigg\}+\frac{9}{8}\frac{1}{au_{b}}\frac{\left(1+(\frac{3}{10})\left(\frac{au_{H}}{au_{b}}\right)^{4}\right)}{(1+(\frac{1}{au_b})^{4})^{2}}\bigg\{30\left(\frac{l}{a}\right)^{5}\nonumber\\
&+&2\left(\frac{d}{a}\right)^{5}+80\left(\frac{l}{a}\right)^{4}\left(\frac{d}{a}\right)+ 40\left(\frac{l}{a}\right)^{2}\left(\frac{d}{a}\right)^{3}+10\left(\frac{l}{a}\right)\left(\frac{d}{a}\right)^{4}\bigg\}\bigg]~.
\end{eqnarray}
Once can see that the mutual information at zero temperature given in eq.(\ref{mi2}) is recovered by setting $u_{H}=0$ in the above result. In this case also, the mutual information is a divergent quantity. All the discussions made earlier in the zero temperature case also hold in this case. 
\section{Conclusion}\label{Sec6}
In this paper we have holographically computed various entanglement measures, namely, entanglement entropy, entropic $c$-function, mutual information and entanglement of purification for noncommutative super Yang-Mills theory in $3+1$-spacetime dimensions. Firstly, we compute the length of the strip like subsystem $\frac{l}{a}$ (by keeping the lengths corresponding to the other directions fixed) by following a systematic analytical approach. We also show that our analytically computed results are in good agreement with that computed numerically. We observe that $\frac{l}{a}$ has a critical length scale $\frac{l_c}{a}$ which primarily points out two distinct domains in the theory, one is for $l<l_c$ and other one is for $l>l_c$. The domain corresponding to $l<l_c$ and $au_{t}>>1,au_{t}\sim au_{b}$ is the deep UV domain of the theory. On the other hand, the domain $l>l_c$ has two solutions depending upon the value of the turning point $au_t$. The domain with $l>l_c,~au_t\ll 1$ is the deep IR domain and the domain with $l>l_c,~au_b>au_t\gg 1$ is the deep noncommutative domain. We then holographically compute the entanglement entropy corresponding to a strip like subsystem $A$. Keeping in mind the presence of the length $l_c$, we have computed the holographic entanglement entropy for both $l>l_{c}$ and $l<l_{c}$ regimes. Once again we observe that our analytical results are in good agreement with that computed numerically. We observe that the divergent part of the HEE is not universal for $l<l_{c}$ since it depends on the subsystem size $l$. For $l>l_c$, we obtain the expected universal divergent part in the HEE (independent of the subsystem size $l$). Our analysis also reveal that extremal surfaces exists for any $l$. In order to probe the signature of noncommutativity on the number of degrees of freedom, we then proceed to compute the entropic $c$-function. By following the proposed definition for the $c$-function in the literature we observe that in the deep IR limit (which corresponds to the commutative YM theory), the $c$-function is obtained to be a constant number which specifies the number of degrees of freedom of the underlying theory. This definition of entropic $c$-function was proposed only for CFTs. However, the full Lorentz symmetry for NC SYM is broken as $SO(3,1) \rightarrow SO(1,1) \times SO(2)$. This motivates us to redefine the $c$-function in such a way so that we can distinctly point out the signature of noncommutativity. Furthermore, we have done this by keeping the leading order correction due to noncommutativity, only for the scenario $l>l_c$. It has been noted that in the deep IR region, the $c$-function of the NC SYM approaches the constant value $C^{sym}$ corresponding to the usual commutative theory of Yang-Mills. We have then investigated the full behaviour of the $c$-function for arbitary subsystem length. We observe that there are discontinuities in the function at the IR, NC and NC, UV junctions.
We then compute another entanglement measure (for mixed states), namely, the entanglement wedge cross-section which holographically probes the entanglement of purification on the basis of $E_P(A,B)=E_W(A,B)$ duality. We compute the EWCS in both the domains $l>l_c$ and $l<l_c$. The effect of noncommutativity has been noted from the graphical representation of the computed results. The critical separation length ($d_c$) between two subsystems (with fixed lengths) at which the holographic mutual information (mutual correlation) vanishes, has also been computed which also probes the connected to disconnected phase tranformation of the RT surface $\Gamma_{AB}^{min}$. However, we have observed this only in the domain $l>l_c$. It is interesting to note that the property $a^2E_W\geq \frac{1}{2}a^2I(A:B)$ holds for the noncommutative gauge theory. We also study the effect of the UV cut-off on the IR results of these quantities.
We then move on to the finite temperature scenario. In this case, we obtain leading order thermal corrections (by considering terms upto $\mathcal{O}(T^4)$) to the information theoretic measures considered in this paper. This we have done in the limit $au_H\ll au_t\ll 1$. This can be interpreted as the low temperature limit. Another unique feature of noncommutative gauge theory which we come across is the divergent piece of the holographic entanglement entropy. We note that in the finite temperature case, the divergent piece of the HEE consists a term which is temperature dependent. In this case also, the notion of a critical length $l_{c}$ is present. Furthermore, in the presence of a finite temperature we note that the value of the length scale $l_c$ depends upon temperature. We then compute the HEE corresponding to the subsystem lengths $l<l_c$ and $l>l_c$. Once again we compute the minimal cross-section of the entanglement wedge, the holographic mutual information and note down the collective effect of finite temperature and noncommutativity on these. Finally, we would like to mention that it will be very interesting to study the holographic subregion complexity and complexity of purification for NC SYM theory. The gravity dual of NC SYM contains a warp factor which one needs to address in the computation of holographic subregion complexity \cite{Gangopadhyay:2020xox}. We leave these as future works in this direction.  
\section*{Acknowledgements}
ARC would like to thank SNBNCBS for the Junior Research Fellowship. AS would like to acknowledge the support by Council of Scientific and Industrial Research (CSIR) for the Senior Research Fellowship. The authors would like to thank the anonymous referee for useful comments.

\bibliographystyle{hephys}  
\bibliography{Reference}
\end{document}